\documentclass[12pt]{article}
\usepackage{epsfig}
\usepackage{hyperref}
\usepackage{color}
\def\beq{\begin{equation}}
\def\eeq{\end{equation}}
\def\bea{\begin{eqnarray}}
\def\eea{\end{eqnarray}}

\def\nl{\nonumber\\}
\def\roughly#1{\mathrel{\raise.3ex\hbox
{$#1$\kern-.75em\lower1ex\hbox{$\sim$}}}}

\def\lesssim{\mathrel{\hbox{\rlap
{\hbox{\lower4pt\hbox{$\sim$}}}\hbox{$<$}}}}
\def\gtrsim{\mathrel{\hbox{\rlap
{\hbox{\lower4pt\hbox{$\sim$}}}\hbox{$>$}}}}

\def\sla#1{\raise.15ex\hbox{$/$}\kern-.57em #1}
\def\bra#1{\left\langle #1\right|}
\def\ket#1{\left| #1\right\rangle}

\newcommand{\ba}{\begin{array}}
\newcommand{\ea}{\end{array}}

\newcommand{\barBdtn}{\bar{B} \to D \tau \bar{\nu}_{\tau}}

\newcommand{\bcln}{b \to c l^- \bar{\nu}_{l}}

\newcommand{\barBstdtn}{\bar{B} \to D^{*} \tau \bar{\nu}_{\tau}}

\def\BDtaunu{\bar{B} \to D \tau^{-} \bar{\nu_\tau}} 
\def\BDlnu{\bar{B} \to D \ell^{-} \bar{\nu_\ell}}
\def\BDstartaunu{\bar{B} \to D^{*} \tau^{-} \bar{\nu_\tau}}
\def\BDstarlnu{\bar{B} \to D^{*} \ell^{-} \bar{\nu_\ell}}

\def\bra#1{\left\langle #1\right|}
\def\ket#1{\left| #1\right\rangle}

\begin{document}
\begin{flushright}
UMISS-HEP-2012-05\\
TIFR/TH/12-23
\end{flushright}
\begin{center}
\bigskip
{\Large \bf \boldmath Diagnosing New Physics in 
$b \to c \, \tau \, \nu_\tau$ decays in the light of the recent BaBar result} \\
\bigskip
\bigskip
{\large 
Alakabha Datta $^{a,}$\footnote{datta@phy.olemiss.edu},
Murugeswaran Duraisamy$^{a,}$\footnote{duraism@phy.olemiss.edu}\\
and Diptimoy Ghosh $^{b,}$\footnote{diptimoyghosh@theory.tifr.res.in}
}
\end{center}
\begin{flushleft}
~~~~~~~~~~~$a$: {\it Department of Physics and Astronomy, 108 Lewis Hall, }\\ 
~~~~~~~~~~~~~~\,{\it University of Mississippi, Oxford, MS 38677-1848, USA}\\
~~~~~~~~~~~$b$: {\it Tata Institute of Fundamental Research}\\ 
~~~~~~~~~~~~~~~{\it Homi Bhabha Road, Colaba, Mumbai 400005, India}\\
\end{flushleft}
\begin{center}
\bigskip (\today)
\vskip0.5cm {\Large Abstract\\} \vskip3truemm
\parbox[t]{\textwidth}
{The BaBar Collaboration has recently reported the measurement of the 
ratio of the branching fractions of 
$\bar{B} \to D(D^{*}) \tau^{-} \bar{\nu_\tau}$ to 
$\bar{B} \to D(D^{*}) \ell^{-} \bar{\nu_\ell}$ which deviates from the 
Standard Model prediction by 2$\sigma$(2.7$\sigma$). This deviation goes 
up to 3.4$\sigma$ level when the two measurements in the $D$ and $D^*$ modes 
are taken together and could indicate new physics. Using an effective Lagrangian for the new physics, we study the implication of
these results and calculate other observables that can shed light on the nature of
the new physics.  We show that  the measurements of the forward-backward asymmetries and the $\tau$ and $D^*$ polarization fractions can be distinguished among the various couplings of the new physics operators.
}
\end{center}
%
\section{Introduction}
The Standard Model (SM) has been extremely successful in furthering the
understanding of the various measurements of Branching Ratios (BR) and 
asymmetries in the quark sector. 
In the quark flavor sector, the B factories, BABAR and Belle, have produced an enormous  quantity of data in the last decade. There is still a lot of data to be analyzed from both experiments. 
The B factories have firmly established the CKM mechanism as the leading order contributor to CP violating phenomena in the flavor sector involving quarks. New physics (NP) effects can add to the leading order term producing deviations from the Standard Model (SM) predictions. 
But even after this incredible success 
the mechanism of Electroweak Symmetry Breaking (EWSB) and the 
origin of masses in the SM still remain very poorly understood. In this 
respect, the second and third generation quarks and leptons are 
quite special because they are comparatively heavier and are 
expected to be relatively more sensitive to  new physics. 
As an example, in certain versions of the two Higgs doublet models (2HDM) the couplings of the new Higgs bosons are proportional to the masses and so new physics effects are more pronounced for the heavier generations.
Moreover, the constraints on new physics  involving, specially the third generation leptons and quarks, are somewhat weaker allowing for larger new physics effects. Interestingly, the branching ratio of $B \to \tau \nu_{\tau}$ shows some tension with the SM  predictions \cite{belletau} and this could indicate NP \cite{Bhattacherjee:2010ju}, possibly coming from an extended scalar or gauge sector. There is also a seeming violation of universality in the tau lepton coupling to the W suggested by the Lep II data which could indicate new physics associated with the third generation lepton \cite{MartinGon}.

If there is NP involving the third generation leptons one can search for it in $B$ decays such as $\BDtaunu$, $\BDstartaunu$ \cite{nierste}.
The semileptonic decays of B meson to the $\tau$ lepton is 
mediated by a $W$ boson in the SM and it is quite well understood 
theoretically. In many models of new physics  this decay gets 
contributions from additional states like new vector bosons or 
new scalar particles. The exclusive decays $\BDtaunu$ 
and $\BDstartaunu$ are important places to look for NP 
because, being three body decays, they offer a host of observables 
in the angular distributions of the final state particles. The 
theoretical uncertainties of the SM predictions have gone down 
significantly in recent years because of the developments in 
heavy-quark effective theory (HQET). The experimental situation 
has also improved a lot since the first observation of the decay 
$\BDstartaunu$ in 2007 by the Belle Collaboration 
\cite{Matyja:2007kt}. After 2007 many improved measurements have 
been reported by both the BaBaR and Belle collaborations and the 
evidence for the decay  $\BDtaunu$ has also been found 
\cite{Aubert:2007dsa,Adachi:2009qg,Bozek:2010xy}. 
Recently, the BaBar collaboration with their full data sample 
of an integrated luminosity 426 fb$^{-1}$ has reported the measurements 
of the quantities \cite{:2012xj}
\begin{eqnarray}
\label{babarnew}
R(D) &=& \frac{BR(\BDtaunu)}
{BR(\BDlnu)}=0.440 \pm 0.058 \pm 0.042\, ,
\nonumber \\
R(D^*) &=& \frac{BR(\BDstartaunu)}
{BR(\BDstarlnu)}=0.332 \pm 0.024 \pm 0.018 \, .
\end{eqnarray}
The SM predictions for $R(D)$ and $R(D^*)$ are 
\cite{:2012xj,Fajfer:2012vx,Sakaki:2012ft}
\begin{eqnarray}
R(D) &=& 0.297 \pm 0.017 \, ,
\nonumber \\
R(D^*) &=& 0.252 \pm 0.003 \,,
\end{eqnarray}
which deviate from the BaBar measurements by 2$\sigma$ and 2.7$\sigma$ 
respectively. The BaBar collaboration themselves reported a 3.4$\sigma$ 
deviation from SM when the two measurements of Eq.~\ref{babarnew} are taken 
together. 

These deviations could be sign of new physics and already certain models of new physics have been considered to explain the data \cite{npbabar}.
In this work, we calculate various observables in $\BDtaunu$ and $\BDstartaunu$ decays
with new physics. We write the most general effective Lagrangian that affect these decays.
The Lagrangian contains two quark and two lepton  scalar, pseudoscalar, vector, axial vector and tensor operators. Considering  subsets of the NP operators at a time, the  coefficient of these operators can be fixed from the BaBar measurements and then one can  study the effect of these operators on the various observables. 

The paper is organized in the following manner. In Sec. 2 we set up our formalism where we introduce the effective Lagrangian for new physics and define the various observables in $\BDtaunu$ and $\BDstartaunu$ decays. We also present the SM predictions for these observables in that section. In Sec. 3 we present the numerical predictions which include constraints on the new physics couplings as well as predictions for the various observables with new physics. Finally, in Sec. 4 we summarize the results of our analysis.
\section{Formalism}
In the presence of new physics (NP), the effective Hamiltonian for the quark-level transition $\bcln$  can be written in the form \cite{ccLag} 
\bea
{\cal{H}}_{eff} &=& \frac{4 G_F V_{cb}}{\sqrt{2}} \Big[ (1 + V_L)\,[\bar{c} \gamma_\mu P_L b] ~ [\bar{l} \gamma^\mu P_L \nu_l] \, +  V_R \, [\bar{c} \gamma^\mu P_R b] ~ [\bar{l} \gamma_\mu P_L \nu_l] \nl && \, + S_L \, [\bar{c} P_L b] \,[\bar{l}  P_L \nu_l] \, +  S_R \, ~[\bar{c} P_R b] \,~ [\bar{l}  P_L \nu_l]  \,  + T_L \, [\bar{c} \sigma^{\mu \nu} P_L b] \,~[\bar{l} \sigma_{\mu \nu} P_L \nu_l]\Big]\,
\eea
where  $G_F = 1.116637 \times 10^{-5} GeV^{-2}$ is the Fermi coupling constant, $V_{cb}$ is the Cabibbo-Koboyashi-Maskawa (CKM) matrix element.  $P_{L,R} = ( 1 \mp \gamma_5)/2$  is the projector of negative/positive chirality, and we use $\sigma_{\mu \nu} = i[\gamma_\mu, \gamma_\nu]/2$. 
We have assumed the neutrinos to be always left chiral.
Further, we do not assume any relation between $b \to u l^- \nu_l$ and $\bcln$ transitions and hence  do not include constraints from
$B \to \tau \nu_{\tau}$. 
The SM  effective Hamiltonian corresponds to $V_L = V_R = S_L = S_R =  T_L = T_R = 0$. In this paper we will ignore the  tensor interactions. With this simplification we write the effective Lagrangian as 
\bea
\label{eq1:Lag}
{\cal{H}}_{eff} &=&  \frac{G_F V_{cb}}{\sqrt{2}}\Big\{
\Big[\bar{c} \gamma_\mu (1-\gamma_5) b  + g_V \bar{c} \gamma_\mu  b + g_A \bar{c} \gamma_\mu \gamma_5 b\Big] \bar{l} \gamma^\mu(1-\gamma_5) \nu_l \nl && +  \Big[g_S\bar{c}  b   + g_P \bar{c} \gamma_5 b\Big] \bar{l} (1-\gamma_5)\nu_l + h.c \Big\}, \
\eea
where $g_{V,A} =  V_R \pm V_L$ and $g_{S,P} =  S_R \pm S_L$.\vspace*{1mm}

We will now consider the two cases:

\begin{itemize}
\item { Case a :  In this case we will set $S_L, S_R=0$ and assume that the NP affects leptons of only the third generation. This scenario could arise from the exchange of a new charged
$W^{\prime}$ boson \cite{dattaneutrino}.
}

\item{ Case b :  
In this case we will set $V_L, V_R=0$ and assume that the NP affects only leptons of the third generation. This scenario could arise in models with extended scalar sector \cite{dattaBnp}.
}

\end{itemize}

The polar angle differential decay distribution in the momentum  transfer squared $q^2$ for the process $\bar{B} \to D^{(*)} l \nu_l$ can be written in the form
\bea
\label{eq2:dd}
\frac{d \Gamma}{dq^2 d\cos{\theta_l}}&=& \frac{|p_{D^{(*)}}| v_l}{256 \pi^3 m^2_B} \sum_{ \rm polarization} |{\cal{M}}(\bar{B} \to D^{(*)} l \nu_l)|^2\,,
\eea
where $v_l = \sqrt{1-m^2_l/q^2}$ and the momentum of the $D^{(*)}$ meson in the B meson rest frame is denoted as $|p_{D^{(*)}}|=\lambda^{1/2}(m^2_B,m^2_{D^{(*)}},q^2)/2 m_B$  with  $\lambda(a,b,c) = a^2 + b^2 + c^2 - 2 (ab + bc + ca)$. 


\subsection{$\BDstartaunu$  angular distribution}
The  full $\BDstartaunu$  angular distribution is given by, 
\bea
\label{eq13:DDRBstdtn}
\frac{d \Gamma^{D^*}}{dq^2 d\cos{\theta_l}}&=& N |p_{D^*}| \Big[ 2 |{\cal{A}}_0|^2  \sin^2{\theta_l} + (|{\cal{A}}_\parallel|^2+ |{\cal{A}}_\perp|^2)  
  (1 + \cos{\theta_l}^2)  -4 Re[{\cal{A}}_\parallel {\cal{A}}^*_\perp] \cos{\theta_l} \nl && + \frac{m^2_\tau}{q^2} \Big( 2 |{\cal{A}}_0 \cos{\theta_l}-{\cal{A}}_{tP}|^2  + (|{\cal{A}}_\parallel|^2+ |{\cal{A}}_\perp|^2)  \sin^2{\theta_l} \Big) \Big]\,,
\eea
where $\theta_l$ is the angle between the $D^*$ meson and the $\tau$ lepton three-momenta in the $q^2$ rest frame, $N = \frac{G^2_F |V_{cb}|^2 q^2}{256 \pi^3 m^2_B} \Big(1-\frac{m_l^2}{q^2}\Big)^2$ and the amplitude $A_{tP}$ is
\bea
\label{eq14:DDRBstAtP}
{\cal{A}}_{tP} &=& \Big({\cal{A}}_t + \frac{\sqrt{q^2}}{m_\tau} {\cal{A}}_P \Big)\,.
\eea

The differential decay rates for the $\tau$ helicities, $\lambda_\tau = \pm 1/2$, 
various transversity amplitudes and the form factors are defined in  appendix \ref{Dstardetails} and \ref{FF} respectively.

\begin{figure}[h!]
\centering
\includegraphics[width=5cm]{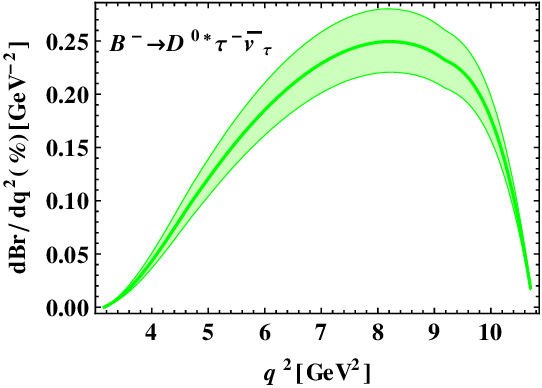}\,
~~\includegraphics[width=5cm]{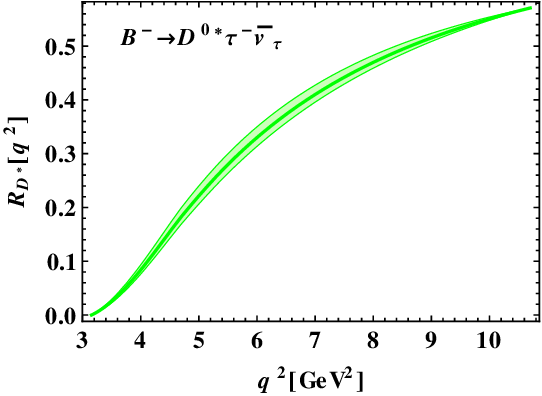}
\caption{The left (right) panels of the figure show the $q^2$ dependence of  DBR ($R_{D^*}(q^2)$ )  for the decay $\bar{B} \to D^{0*} \tau^- \bar{\nu}_\tau$. The bands correspond to uncertainties in $h_{A_1}(1) |V_{cb}|$, $ \rho^2$ and  $R_{0,1,2}(1)$. The errors are added in quadrature. 
\label{fig-DstdBrRDst}}
\end{figure}

The angular distribution allows us to define several observables \cite{Fajfer:2012vx, Sakaki:2012ft}. The starting point is to obtain the decay rates $d\Gamma/dq^2$  for the $\tau$ helicities, $\lambda_\tau = \pm 1/2$ , after performing integration over $\cos{\theta_l}$ , 

\bea
\label{eq15:DDRBstdsqq}
\frac{d \Gamma^{D^*}[\lambda_\tau = -1/2]}{dq^2 }&=& \frac{8  N |p_D|}{3} ~~|{\cal{A}}_T|^2\,,\nl
\frac{d \Gamma^{D^*}[\lambda_\tau = 1/2]}{dq^2}&=& \frac{4  N |p_D| }{3}~~ \frac{m^2_\tau}{q^2}\Big[ |{\cal{A}}_T|^2 +  3 | {\cal{A}}_{tP}|^2   \Big] \,,
\eea
where $|{\cal{A}}_T|^2 = |{\cal{A}}_0|^2 + |{\cal{A}}_\parallel|^2+ |{\cal{A}}_\perp|^2 $. The summation of these  rates give the total differential branching ratio (DBR):
 \bea
\label{eq16:DDRBstdsqq2}
\frac{d Br[\BDstartaunu]}{dq^2 }&=&  \frac{8 N |p_D|\tau_{B} }{3 } ~~ \Big[|{\cal{A}}_T|^2\Big(1 + \frac{m^2_\tau}{2q^2}\Big) + \frac{3 m^2_\tau}{2q^2}| {\cal{A}}_{tP}|^2   \Big]\,.
\eea
where $\tau_{B}$ is the B meson life-time. Furthermore, one can also explore the $q^2$ dependent ratio
\bea
\label{eq20:DDRRDst2}
R_{D^*}(q^2) &=&\frac{d Br[\BDstartaunu]/dq^2 }{d Br[\BDstarlnu]/dq^2}\,,
\eea
where $l$ denotes the light lepton $(e, \mu)$. The ratio  $R_{D^*}(q^2)$ is independent of the  form factor $h_{A_1}(w)$. The SM predictions  of DBR and $R_{D^*}(q^2)$ for  the   decay $\bar{B} \to D^{0*} \tau^- \bar{\nu}_\tau$  are shown in Fig.~\ref{fig-DstdBrRDst}.  The numerical values of  $B \to D^*$ form factor parameters $h_{A_1}(1) |V_{cb}|$, $ \rho^2$ and  $R_{0,1,2}(1)$ are given in appendix \ref{FF}. In the SM,   the DBR for the decay $\bar{B} \to D^{0*} \tau^- \bar{\nu}_\tau$ peaks ($\approx 0.27 \%$ $\mathrm{GeV^{-2}}$) at $q^2\approx 8 $ $\mathrm{GeV^2}$, and the ratio  $R_{D^*}(q^2)$ rises to more than $ 50 \%$ at large $q^2$. 
It is clear from the plot that the uncertainty in $R_{D^*}(q^2)$ is less than that in the DBR.

Next, we define the forward-backward asymmetry (AFB) in the angular distribution  by integrating over $\cos{\theta_l}$ as
 \bea
\label{eq18:DDRAFst}
[A_{FB}]_{D^*}(q^2)=\frac{(\int^1_0- \int^0_{-1}) d\cos{\theta_l} \frac{d \Gamma^{D^*}}{dq^2 d\cos{\theta_l}} }{ \frac{d \Gamma^{D^*}}{dq^2 }}= - \frac{3 }{2} \frac{ \Big(Re[{\cal{A}}_\parallel {\cal{A}}^*_\perp] +\frac{ m_\tau^2}{q^2} Re[{\cal{A}}_0 {\cal{A}}^*_{tP}]\Big)}{|{\cal{A}}_T|^2\Big(1 + \frac{m^2_\tau}{2q^2}\Big) + \frac{3 m^2_\tau}{2q^2}| {\cal{A}}_{tP}|^2}\,.
\eea 
The perpendicular transversity amplitude ${\cal{A}}_{\perp} $ is proportional to $ \sqrt{w^2-1}$ (see appendix \ref{Dstardetails} for the details), hence for the light leptons  $[A_{FB}]_{D^*}(q^2)$ vanishes at the end-points due to the kinematics. One can obtain the angular distribution only for the transversely polarized $D^*$ meson from Eq.~(\ref{eq13:DDRBstdtn}) by dropping  the amplitudes ${\cal{A}}_0$ and ${\cal{A}}_{tP}$.
We now define the forward-backward asymmetry for the transversely polarized $D^*$ meson by integrating over $\cos{\theta_l}$ as in Eq.~(\ref{eq18:DDRAFst}) \cite{Neubert:1993mb}:
\bea 
\label{eq18:DDRAFstDstar}
[A^T_{FB}]_{D^*}(q^2) &=& - \frac{3 }{2} \frac{ Re[{\cal{A}}_\parallel {\cal{A}}^*_\perp] }{(|{\cal{A}}_\parallel|^2+ |{\cal{A}}_\perp|^2) \Big(1 + \frac{m^2_\tau}{2q^2}\Big)}\,.
\eea 

\begin{figure}[h!]
\centering
\includegraphics[width=5cm]{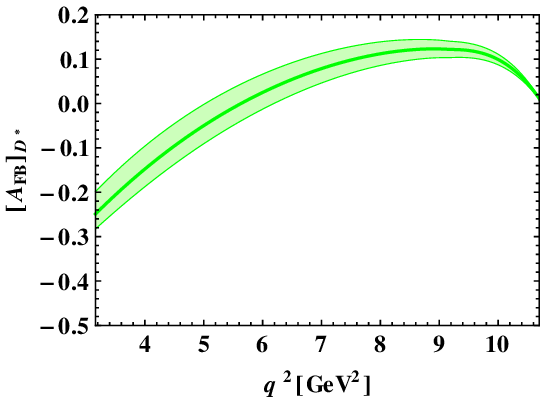}\,
~~\includegraphics[width=5cm]{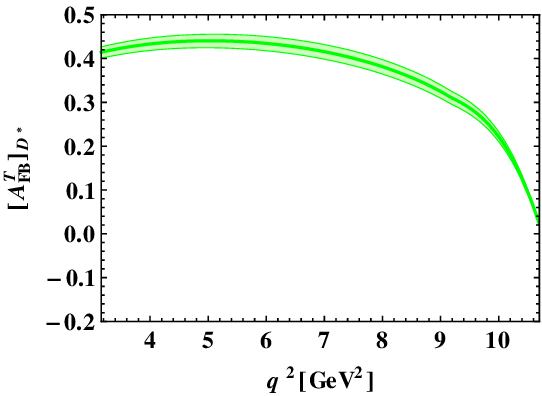}
\caption{The left (right) panel shows the $q^2$ dependence of $[A_{FB}]_{D^*}$ ($[A^{T}_{FB}]_{D^*}$ )  for the decay $\bar{B} \to D^{0*} \tau^- \bar{\nu}_\tau$.  The bands correspond to uncertainties in $ \rho^2$ and  $R_{0,1,2}(1)$. The errors are added in quadrature.
\label{fig-AFBDstSM}}
\end{figure}

Fig.~\ref{fig-AFBDstSM} shows the SM predictions for $[A_{FB}]_{D^*}(q^2)$ and $[A^{T}_{FB}]_{D^*}(q^2)$ in  the   decay $\BDstartaunu$.  $[A_{FB}]_{D^*}(q^2)$ is $\sim 20 \%$ and negative at low $q^2$, and has a zero crossing at $q^2 \approx 5.6 4 $ $\mathrm{GeV}^2$. However, the asymmetry  $[A^{T}_{FB}]_{D^*}(q^2)$ is always positive and  large ($\sim 40\% $) at low $q^2$.

We now define the longitudinal and transverse polarization fractions of the $D^*$ meson from Eq.~\ref{eq16:DDRBstdsqq2} as
\bea
\label{eq17:DDRBstfLDst}
F^{D^*}_L &=& \frac{ |{\cal{A}}_0|^2 (1 + \frac{m^2_\tau}{2q^2}\Big) + \frac{3 m^2_\tau}{2q^2}| {\cal{A}}_{tP}|^2}{|{\cal{A}}_T|^2\Big(1 + \frac{m^2_\tau}{2q^2}\Big) + \frac{3 m^2_\tau}{2q^2}| {\cal{A}}_{tP}|^2}\,,\nl
F^{D^*}_T &=& \frac{ (|{\cal{A}}_\parallel|^2+ |{\cal{A}}_\perp|^2) (1 + \frac{m^2_\tau}{2q^2}\Big)}{|{\cal{A}}_T|^2\Big(1 + \frac{m^2_\tau}{2q^2}\Big) + \frac{3 m^2_\tau}{2q^2}| {\cal{A}}_{tP}|^2}\,,
\eea
where $F^{D^*}_L + F^{D^*}_T = 1$. The $D^*$ polarization fractions can be measured
by fitting to  the decay distribution in Eq.~\ref{eq13:DDRBstdtn} or from $D^*$ decays.

Finally, one can also define the  longitudinal polarization fraction of the $\tau$ lepton  in the  $q^2$ rest frame as
 \bea
\label{eq16:DDRBstfLtau}
P^{* \tau}_L (q^2)&= & \frac{\frac{d \Gamma^{D^*}[\lambda_\tau =- 1/2]}{dq^2 }-\frac{d \Gamma^{D^*}[\lambda_\tau = 1/2]}{dq^2}}{\frac{d \Gamma^{D^*}}{dq^2 }} = \frac{|{\cal{A}}_T|^2\Big(1- \frac{m^2_\tau}{2q^2}\Big)- \frac{3 m^2_\tau}{2q^2}| {\cal{A}}_{tP}|^2  }{|{\cal{A}}_T|^2\Big(1 + \frac{m^2_\tau}{2q^2}\Big) + \frac{3 m^2_\tau}{2q^2}| {\cal{A}}_{tP}|^2   }\,.
\eea
The $\tau$ polarization can be measured from the decays of the $\tau$.

\begin{figure}[h!]
\centering
\includegraphics[width= 5cm]{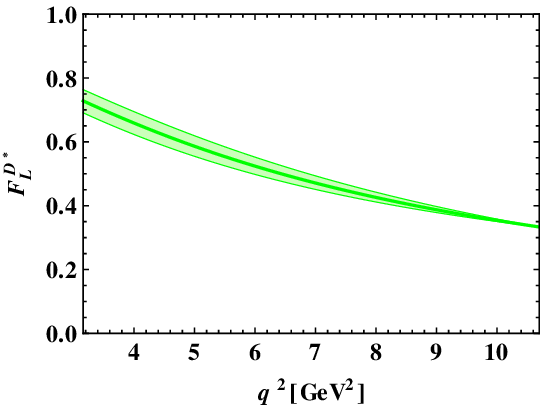}\,
~~\includegraphics[width=5cm]{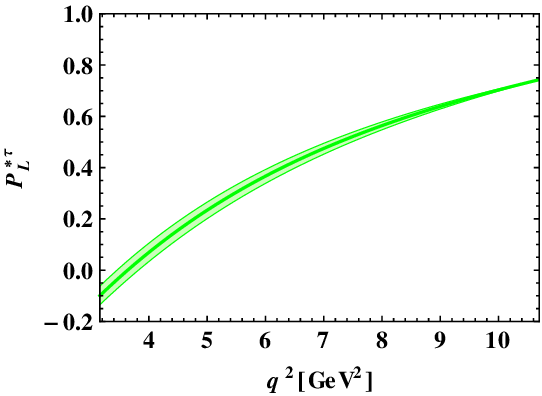}
\caption{The left (right) panel shows the $q^2$ dependence of  $F^{D^*}_L$ ($P^{* \tau}_L$) for the decay $\bar{B} \to D^{0*} \tau^- \bar{\nu}_\tau$.  The bands correspond to uncertainties in $ \rho^2$ and  $R_{0,1,2}(1)$. The errors are added in quadrature.
\label{fig:polDstatuSM}}
\end{figure}

The polarization fractions are independent of the  form factor $h_{A_1}(w)$. The SM predictions of the longitudinal  polarization fractions of $D^*$ and  $\tau$  are shown in Fig.~\ref{fig:polDstatuSM}. In the SM, $F^{D^*}_L(q^2)$ can be as large as 0.75 at low $q^2$, and it decreases to about 0.4 at high $q^2$. On the other hand, $P^{*\tau}_L(q^2)$ is about -0.18 at very low $q^2$ and has a zero-crossing at $q^2 \approx 3.64 $  $\mathrm{GeV}^2$ and increases to about 0.7 at high $q^2$.

\subsection{$\BDtaunu$  angular distribution}
\label{barBdtn}
The full $\BDtaunu$  angular distribution can be written as, 
\bea
\label{eq5:DDRBdtn}
\frac{d \Gamma^D}{dq^2 d\cos{\theta_l}}&=& N |p_D|~ \Big[ 2 |H_0|^2  \sin^2{\theta_l}  + 2 \frac{m^2_\tau}{q^2} (H_0 \cos{\theta_l}-H_{tS})^2 \Big]\,,
\eea
where
\bea
H_{tS} &=& \Big(H_t - \frac{\sqrt{q^2}}{m_\tau} H_S \Big)\,.
\eea
The differential decay rates for the  $\tau$ helicities, $\lambda_\tau = \pm 1/2$, and helicity amplitudes  $H_0$ are defined in appendix \ref{Ddetails}.

As in the previous section, we can define several observables using the $\BDtaunu$ 
  angular distribution \cite{ Sakaki:2012ft}. The starting point is to obtain the decay rates $d\Gamma/dq^2$  for the $\tau$ helicities, $\lambda_\tau = \pm 1/2$. After performing integration over $\cos{\theta_l}$, one can obtain: 
\bea
\label{eq6:DDRBdtn2}
\frac{d \Gamma^D[\lambda_\tau = -1/2]}{dq^2 }&=& \frac{8}{3} N |p_D|  |H_0|^2 \,,\nl
\frac{d \Gamma^D[\lambda_\tau = 1/2]}{dq^2}&=& \frac{4}{3} N |p_D|  \frac{m^2_\tau}{q^2} ( |H_0|^2 + 3 | H_{tS}|^2) \,,
\eea
 and the summation of these differential decay rates give the DBR
 \bea
\label{eq7:DDRBdtn3}
\frac{d Br[\BDtaunu]}{dq^2 }&=& \frac{8 N |p_D| \tau_B}{3 \hbar}  \Big[ |H_0|^2 (1+  \frac{m^2_\tau}{2q^2}) +   \frac{3 m^2_\tau}{2q^2} | H_{tS}|^2\Big]\,.
\eea

As in the previous section, one can also explore the $q^2$ dependent ratio
\bea
\label{eq20:DDRBdtn3}
R_{D}(q^2) &=&\frac{d Br[\BDtaunu]/dq^2 }{d Br[\BDlnu]/dq^2}\,.
\eea
The ratio  $R_{D}(q^2)$ is independent of the form factors.
The SM predictions  of DBR and $R_{D}(q^2)$ for  the   decay $\bar{B} \to D^0 \tau^- \bar{\nu}_\tau$  are shown in Fig.~\ref{fig:DdBrRD}.  The numerical values of the free parameters  in the $B \to D$  form factors  are given in  appendix \ref{Ddetails}. In the SM, the DBR for the decay $\bar{B} \to D^0 \tau^- \bar{\nu}_\tau$ peaks ($\approx 0.14 \%  ~\mathrm{GeV}^{-2}$ ) at  $q^2\approx 7 $ $\mathrm{GeV^2}$, while $R_{D}(q^2)$ shows almost a linear behavior with $q^2$. 

\begin{figure}[h!]
\centering
\includegraphics[width=5cm]{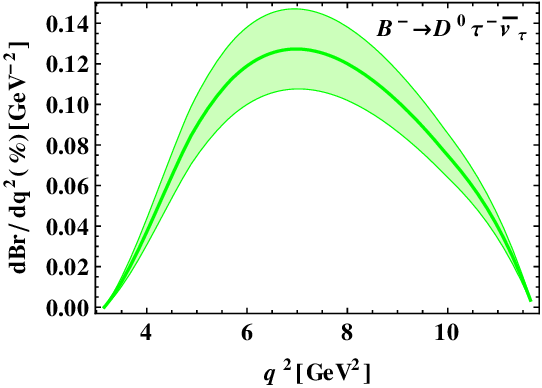}\,
~~\includegraphics[width=5cm]{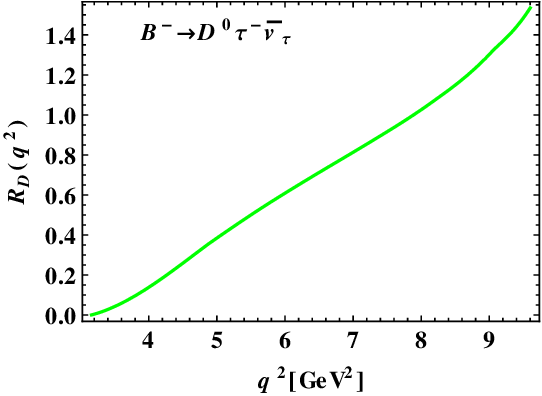}
\caption{The left (right) panel shows the $q^2$ dependence of DBR ( $R_{D}(q^2)$ )  for the decay $\bar{B} \to D^0 \tau^- \bar{\nu}_\tau$. The bands correspond to uncertainties in $V_{1}(1) |V_{cb}|$, $ \rho^2$ and  $\rho^2_1$. The errors are added in quadrature.
\label{fig:DdBrRD}}
\end{figure}

Next, we define the forward-backward asymmetry in the angular distribution by integrating over  $\cos{\theta_l}$ as
\bea
\label{eq18:AFBD}
[A_{FB}]_D(q^2)= \frac{(\int^0_{-1}-\int^1_0 ) d\cos{\theta_l} \frac{d \Gamma^D}{dq^2 d\cos{\theta_l}} }{ \frac{d \Gamma^D}{dq^2 }}=
\frac{3 m_\tau^2}{2q^2} \frac{ Re[H_0 H^*_{tS}]}{|H_0|^2 (1+  \frac{m^2_\tau}{2q^2}) +   \frac{3 m^2_\tau}{2q^2} | H_{tS}|^2}\,.
\eea 
Finally, we define the   longitudinal polarization fraction of $\tau $ in the  $q^2$ rest frame as
 \bea
\label{eq8:DDRfL}
P^{ \tau}_L(q^2)&= & \frac{\frac{d \Gamma^D[\lambda_\tau = 1/2]}{dq^2 }-\frac{d \Gamma^D[\lambda_\tau = -1/2]}{dq^2}}{\frac{d \Gamma^D}{dq^2 }}
= \frac{|H_0|^2 (  \frac{m^2_\tau}{2q^2}-1) +   \frac{3 m^2_\tau}{2q^2} | H_{tS}|^2}{|H_0|^2 (1+  \frac{m^2_\tau}{2q^2}) +   \frac{3 m^2_\tau}{2q^2} | H_{tS}|^2}\,.
\eea \vspace*{1mm}
The $\tau$ polarization can be measured from the decays of the $\tau$.

\begin{figure}[h!]
\centering
\includegraphics[width=5cm]{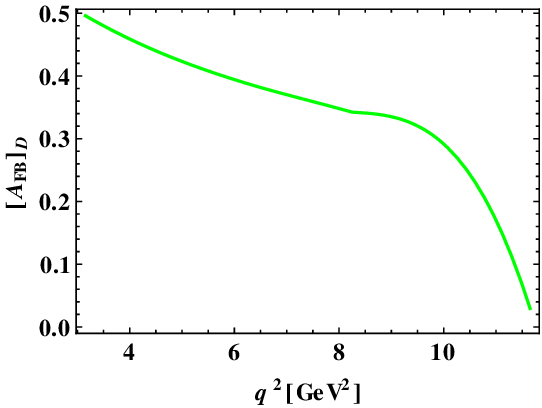}\,,\includegraphics[width=5cm]{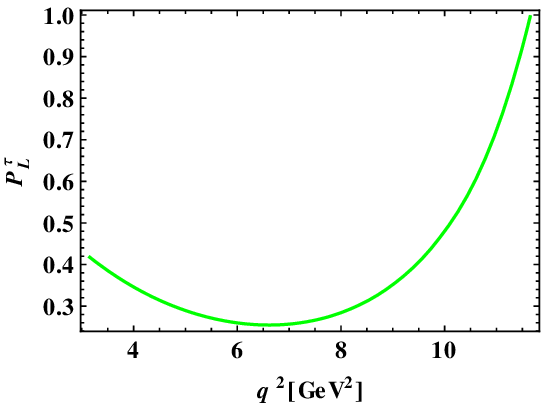}
\caption{The left (right) panel shows the $q^2$ dependence of $[A_{FB}]_{D}$ ($P^{ \tau}_L)$ for the decay $\bar{B} \to D^{0} \tau^- \bar{\nu}_\tau$.
\label{fig:AFBDSM}}
\end{figure}

 Fig.~\ref{fig:AFBDSM} shows the SM predictions for $[A_{FB}]_{D}(q^2)$ and 
$P^{ \tau}_L(q^2)$ for the decay $\bar{B} \to D^{0} \tau^- \bar{\nu}_\tau$. 
The forward-backward asymmetry, $[A_{FB}]_{D}(q^2)$, is about 
$ \sim 50 \%$ at low $q^2$ and decreases  with increasing $q^2$. 
The $\tau$ polarization, $P^{\tau}_L$, is about 0.4 at low $q^2$  and starts to increase  after  $q^2 = 8 \mathrm{GeV^2} $.

In the next section we shall study the effect of the new physics couplings $V_{L,R}$, and  $S_{L,R}$ 
on the above observables.

\section{Numerical analysis with NP}
In the numerical analysis, as indicated earlier, we consider  two cases to study the   new physics effects on DBR, the ratios $R_{D(D^{*})}(q^2)$, the forward-backward  asymmetries, and polarization fractions. In the first case we consider only vector/axial-vector NP couplings while in the second case we consider only scalar/pseudoscalar NP couplings. 
The numerical values of $B \to D$ and $B \to D^*$ form factors in the heavy quark effective theory framework are summarized in appendix \ref{FF}. A detail discussion of these form factors can be found in \cite{Caprini:1997mu}. 

In our numerical analysis, we constrain  both complex/real NP couplings $V_{L,R}$ and $S_{L,R}$ using the measured $R(D)$ and $R(D^*)$ in Eq.~(\ref{babarnew}) at 95\% C.L. We also vary the free parameters in the form factors discussed in appendix \ref{FF} within their error bars. All the other numerical values are taken from \cite{Nakamura:2010zzi} and \cite{Asner:2010qj}. The allowed  ranges for  NP couplings are then used for predicting the allowed ranges  for the observables discussed earlier.

\subsection{$\bar{B} \to D^{0*} \tau^- \bar{\nu}_\tau$}
\subsubsection{Pure $V_L$ and $V_R$ couplings present}
\begin{figure}[h!]
\centering
\includegraphics[width=4.7cm,height=4.7cm]{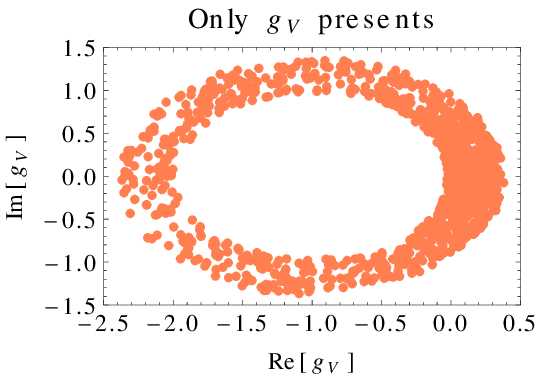}\,
\includegraphics[width=4.7cm,height=4.7cm]{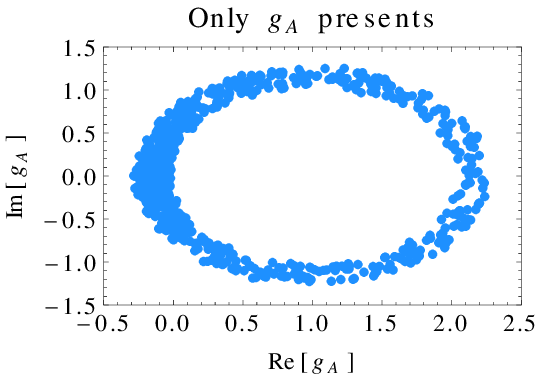}\,
\includegraphics[width=4.7cm,height=4.35cm]{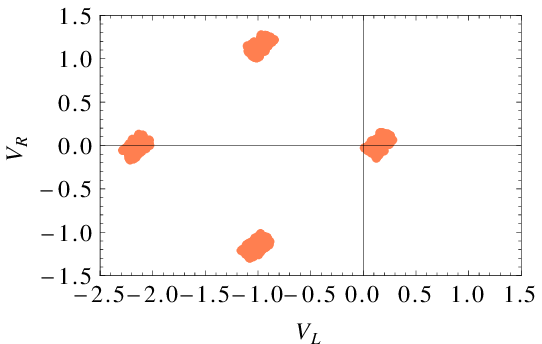}
\caption{The constraints on the complex coupling $g_V = V_R + V_L$, $g_A = V_R - V_L$ and the real
$(V_R, V_L)$ couplings  at 95\% C.L. 
\label{fig:OnlygVgA}}
\end{figure}

The combination of the couplings $g_V = V_R + V_L$ appears  in both $R(D)$ and $R(D^*)$, while $g_A = V_R - V_L$  appears only in $R(D^*)$. $V_R$ and $V_L$ receive constraints from both  $R(D)$ and $ R(D^*)$.
If new physics is established in both $R(D)$ and $ R(D^*)$ then the case of pure $g_A$ coupling is ruled out.
The constraints on the complex couplings $g_V$  and $g_A$  are shown in the colored region of Fig.~\ref{fig:OnlygVgA} (left) and (center). Fig.~(\ref{fig:OnlygVgA})(right) shows the constraints on the real couplings $V_L$ and $V_R$. The real couplings  are severely constraints by  the recent $R(D)$ and $R(D^*)$ measurements. The plot indicates that the data prefers either pure vector or pure axial vector couplings. Henceforth we will consider all three cases which include pure $g_V$ and $g_A$ complex couplings and real $(V_L,V_R)$ couplings.. 

\subsubsection{Only $g_V$ coupling present}
In this section we consider only  vector coupling $g_V=V_R+V_L$.
The NP vector coupling $g_V$ appears only in the amplitude ${\cal{A}}_\perp$ and except for the forward-backward asymmetries, it does not significantly affect any other observables discussed earlier for the decay $\bar{B} \to D^{0*} \tau^- \bar{\nu}_\tau$. 
\begin{figure}[h!]
\centering
\includegraphics[width=5.5cm]{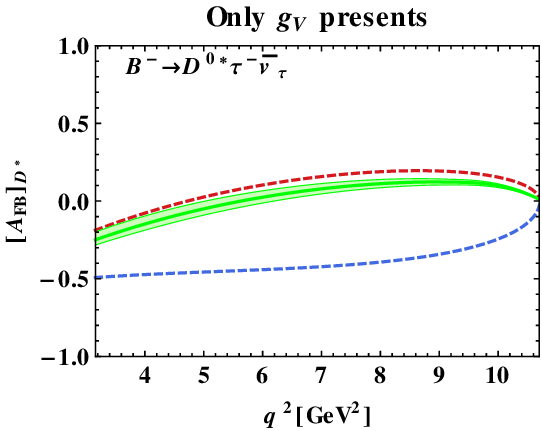}\,~~\includegraphics[width=5.5cm]{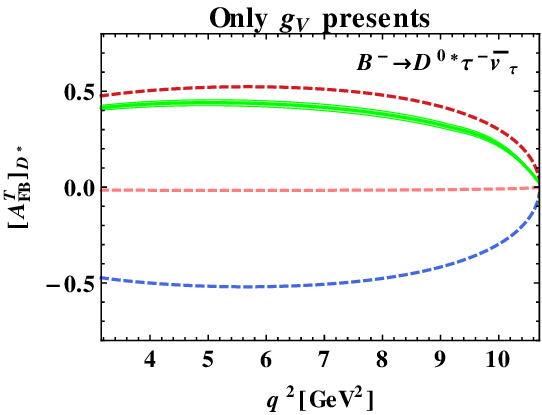}
\caption{The left (right) panel shows the $q^2$ dependence of $[A_{FB}]_{D^*}$ ($[A^{T}_{FB}]_{D^*}$)  for the decay $\bar{B} \to D^{0*} \tau^- \bar{\nu}_\tau$.  The dashed lines show predictions for some  representative values of $g_V$. For example blue lines  correspond to $g _V = 2.34 e^{-i 3.08}$. }
\label{fig:AFBDstonlygV}
\end{figure}
One can see from 
Fig.~\ref{fig:AFBDstonlygV} that the coupling $g_V$ can enhance the magnitude of  $[A_{FB}]_{D^{*} }$ up to $50\%$ at low $q^2$ and it can have different zero-crossing point than the SM. In the SM, no zero-crossing is allowed for $[A^T_{FB}]_{D^{*} }$, however in the presence of $g_V$, $[A^T_{FB}]_{D^{*} }$ may have zero-crossing. Also, $[A^T_{FB}]_{D^{*} }$ can reach up to $50\%$ at low $q^2$, and it can have either positive or negative sign.

\subsubsection{Only $g_A$ coupling present}
\begin{figure}[h!]
\centering
\includegraphics[width=5.5cm]{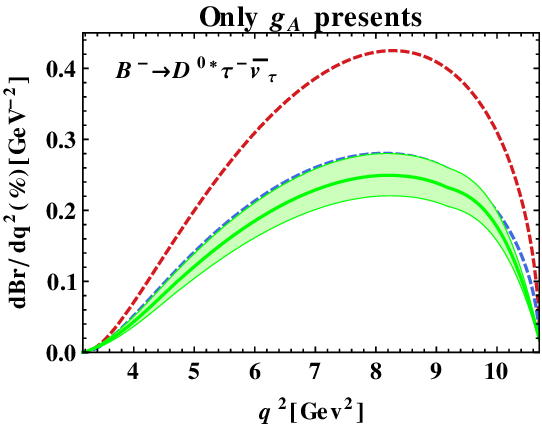}\,~~\includegraphics[width=5.5cm]{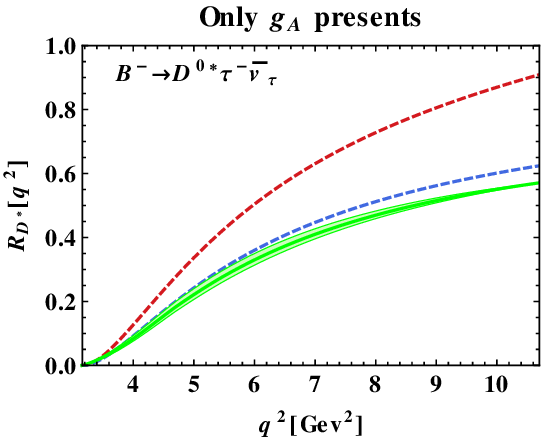}
\caption{The left (right) panel shows the $q^2$ dependence of DBR ($R_{D^*}(q^2)$)  for the decay $\bar{B} \to D^{0*} \tau^- \bar{\nu}_\tau$.  The dashed lines show predictions for some  representative values of  $g_A$. For example, the red lines correspond to $g _A = 0.31 e^{-i 2.62}$.}
\label{fig:DstdBrRDstonlygA}
\end{figure}

\begin{figure}[h!]
\centering
\includegraphics[width=5.5cm]{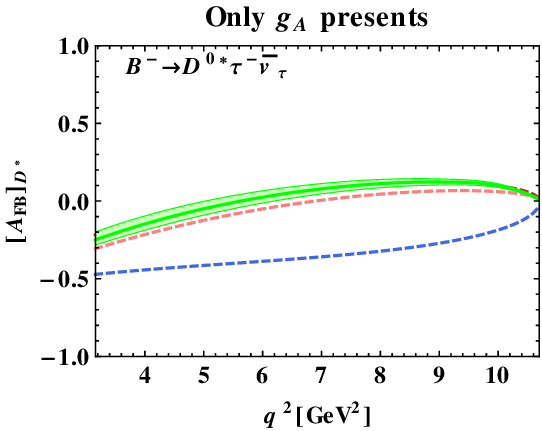}\,~~\includegraphics[width=5.5cm]{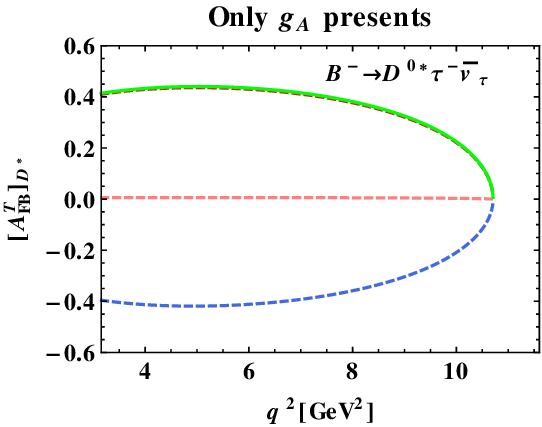}
\caption{The left (right) panels of the figure show the $q^2$ dependence of $[A_{FB}]_{D^*}$ ($[A^{T}_{FB}]_{D^*}$) for the decay $\bar{B} \to D^{0*} \tau^- \bar{\nu}_\tau$.  The dashed lines show predictions for some  representative values of $g_A$. For example the blue lines  correspond to $g _A = 2.34 e^{-i 3.08}$. }
\label{fig:AFBDstonlygA}
\end{figure}

In this section, we consider only pure axial vector coupling $g_V=V_R-V_L$.
In this case, except  ${\cal{A}}_\perp$  all other amplitudes  depend on the new axial-vector coupling $g_A$ while the amplitude ${\cal{A}}_{P}$ is zero. Thus,  the  coupling $g_A$ does not significantly change the values for the polarization fractions of the  $D^*$ meson and the $\tau$ lepton from their SM predictions. In Fig.~\ref{fig:DstdBrRDstonlygA} we show the DBR and $R_{D^*}(q^2)$ in the presence of  $g_A$. The coupling $g_A$ can enhance DBR up to $ 0.4\%$ $\rm GeV^{-2}$ at $q^2\approx 8.5 \mathrm{GeV^2}$, and $R_{D^*}(q^2)$ can be as high as about 0.9 at high $q^2$.\vspace*{1mm}
As shown in Fig.~\ref{fig:AFBDstonlygA}, the  coupling $g_A$ can enhance the magnitude of $[A_{FB}]_{D^{*} }$ to about $50\%$ at low $q^2$ and it can now have different zero-crossing  point than the SM. In the SM, no zero-crossing is allowed for $[A^T_{FB}]_{D^{*} }$, however in the presence of $g_A$, $[A^T_{FB}]_{D^{*} }$ may have zero-crossing. Also, $[A^T_{FB}]_{D^{*} }$ can take either positive or negative sign. The results are similar to the case where only $g_V$ coupling is present.

\subsubsection{Both $V_{L,R}$ coupling are present and are real}
\begin{figure}[h!]
\centering
\includegraphics[width=5.5cm]{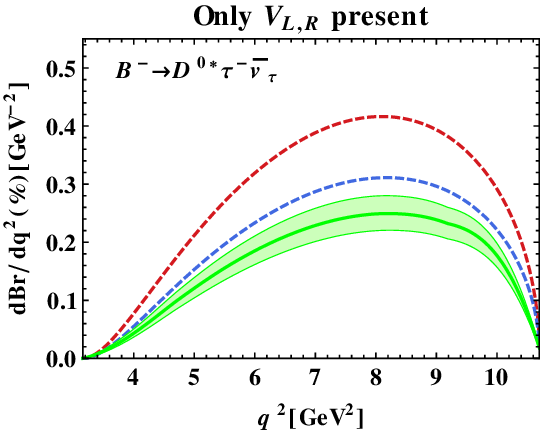}\,~~
\includegraphics[width=5.5cm]{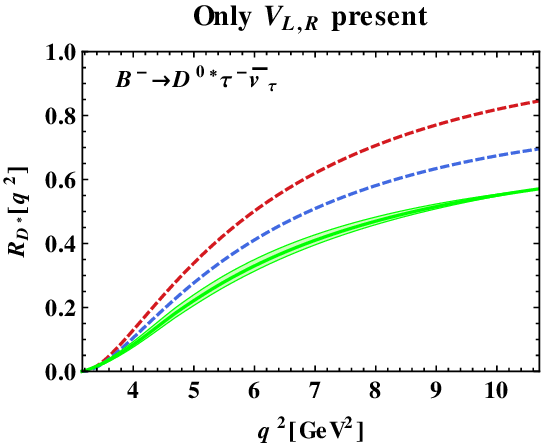}
\caption{The left (right) panel shows the $q^2$ dependence of DBR ($R_{D^*}(q^2)$) for the decay $\bar{B} \to D^{0*} \tau^- \bar{\nu}_\tau$.  The dashed lines show predictions for some  representative values of $(V_L,V_R)$. For example the red lines in  correspond to  $(V_L,V_R) = (-0.97, 1.24)$.
\label{fig:DstdBrRDstonlyLR}}
\end{figure}

Finally, we consider the case where both $V_{L,R}$ coupling are present and are real.
In Fig.~\ref{fig:DstdBrRDstonlyLR} we show the DBR and $R_{D^*}(q^2)$ in the presence of both $V_L$ and $V_R$ real couplings.  These couplings can enhance the DBR upto $ 0.4\%$ $\rm GeV^{-2}$ at $q^2\approx 8.5 \mathrm{GeV^2}$, and $R_{D^*}(q^2)$ can be increased to about 0.9 at high $q^2$.\vspace*{1mm}

\begin{figure}[h!]
\centering
\includegraphics[width=5.5cm]{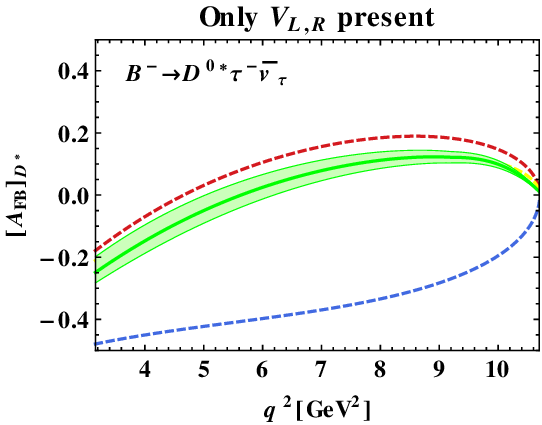}\,
~~\includegraphics[width=5.5cm]{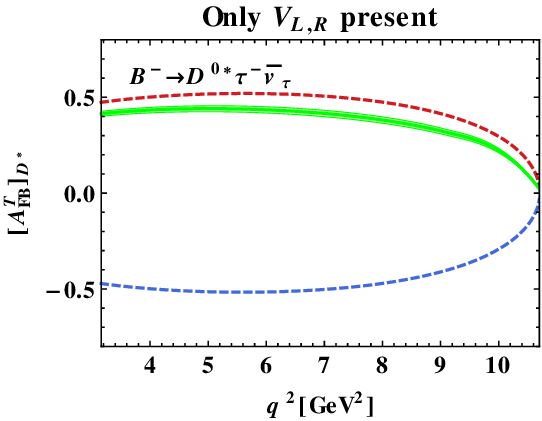}
\caption{The left (right) panel shows the $q^2$ dependence of $[A_{FB}]_{D^*}$ ($[A^{T}_{FB}]_{D^*}$) for the decay $\bar{B} \to D^{0*} \tau^- \bar{\nu}_\tau$.  The dashed lines show predictions for some  representative values of  $g_V$. For example the blue lines  correspond to $(V_L,V_R) = (-1.02, 1.08)$ in the left panel and $(V_L,V_R) = (-0.85, 1.21)$ in the  right panel. 
\label{fig:AFBDstonlyLR}}
\end{figure}

One can see from Fig.~\ref{fig:AFBDstonlyLR}, the  couplings $V_L$ and $V_R$  can negatively enhance $[A_{FB}]_{D^{*} }$ upto $50\%$ at low $q^2$ and it can have a different zero-crossing than  the SM. In the SM, no zero-crossing is allowed for $[A^T_{FB}]_{D^{*} }$, however in the presence of these new couplings $[A^T_{FB}]_{D^{*} }$ may have zero-crossing. Also, $[A^T_{FB}]_{D^{*} }$ can reach upto $50\%$ at low $q^2$, and it can take either positive or negative values.
  
The polarization fractions of the  $D^*$ meson are almost independent of the new couplings  $V_L$ and $V_R$. The tau lepton polarization fraction too does not depend on $V_L$ and $V_R$.

\subsubsection{Pure $S_L$ and $S_R$ couplings present}
In this section we consider the scalar and pseudo-scalar couplings $S_{L,R}$.
The combination of the couplings $S_R + S_L$ appears  only  in $R(D)$, while  $S_R - S_L$  appears only in $R(D^*)$. 
If new physics is established in both $R(D)$ and $ R(D^*)$ then the cases of pure
$S_R \pm S_L$  couplings are ruled out.

\begin{figure}[h!]
\centering
\includegraphics[width=4.7cm,height=4.7cm]{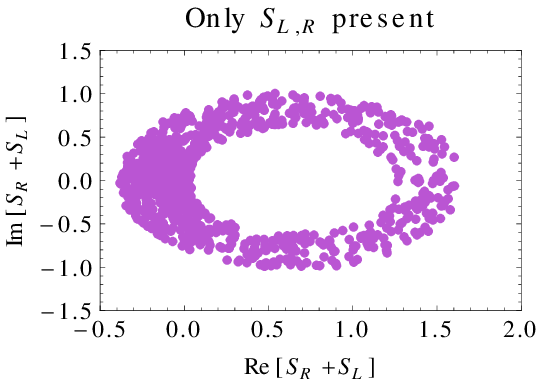}\,
\includegraphics[width=4.7cm,height=4.5cm]{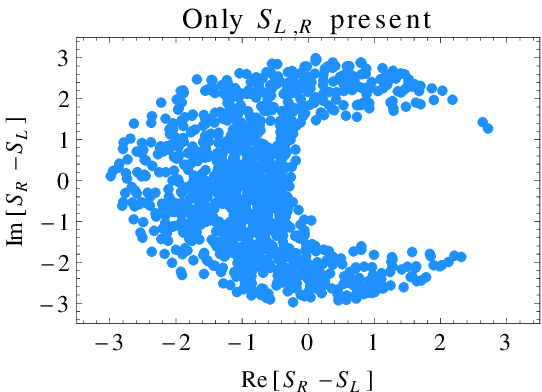}
\includegraphics[width=4.7cm,height=4.2cm]{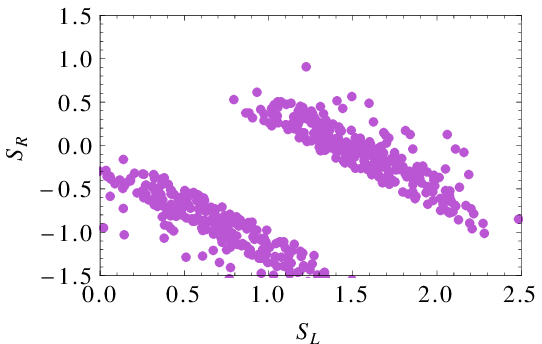}
\caption{The constraints on the couplings for complex ($S_R + S_L$) (left), complex ($S_R - S_L$) (center), and real $ S_L$ and $ S_R$ (right) at 95\% C.L.~. 
\label{fig:OnlygSgP}}
\end{figure}

Hence, $S_R$ and $S_L$ get constrained from both $R(D)$ and $ R(D^*)$. The constraints on the complex couplings $S_R + S_L$ and $S_R - S_L$  are shown 
in Fig.~\ref{fig:OnlygSgP} (left and middle). Fig.~\ref{fig:OnlygSgP}(right) shows the constraints on the real couplings $S_L$ and $S_R$. The real couplings  are severely 
constrained by  the recent $R(D)$ and $R(D^*)$ measurements though the constraints are relatively weaker than those in the $(V_L, V_R)$ case. 
To simplify our discussion we will take $S_{L,R}$ to be real.
\vspace*{1mm}

\begin{figure}[h!]
\centering
\includegraphics[width=5.5cm]{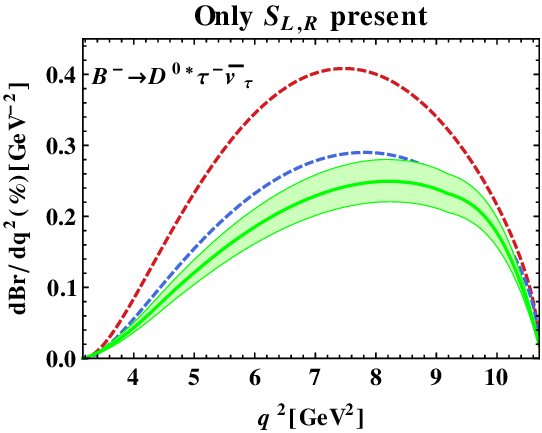}\,
~~\includegraphics[width=5.5cm]{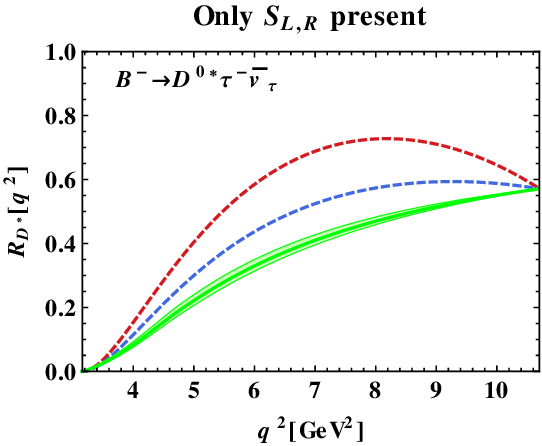}
\caption{The left (right) panel shows the $q^2$ dependence of DBR ($R_{D^*}(q^2)$ ) for the decay $\bar{B} \to D^{0*} \tau^- \bar{\nu}_\tau$.  The dashed lines show predictions for some  representative values of   $S_L$ and $S_R$. For example the red lines  correspond to $(S_L,S_R) = (1.11, -1.30)$. 
\label{fig:DstdBrRDstonlygP}}
\end{figure}

The combination of couplings $S_R-S_L$ appears only in the amplitude ${\cal{A}}_{P}$. In Fig.~\ref{fig:DstdBrRDstonlygP} we show the DBR and $R_{D^*}(q^2)$ in the presence of $S_L$ and $S_R$ couplings. These couplings can  enhance the DBR up to $ 0.4\%$ $\rm GeV^{-2}$ at  $q^2\approx 7.5 \mathrm{GeV^2}$. Note that the peak of the DBR is shifted to low $q^2$ direction relative to the SM.  The ratio, $R_{D^*}(q^2)$, can take the value of  about 0.7 at $q^2\approx 7.5 \mathrm{GeV^2}$.\vspace*{1mm}

\begin{figure}[h!]
\centering
\includegraphics[width=4.5cm]{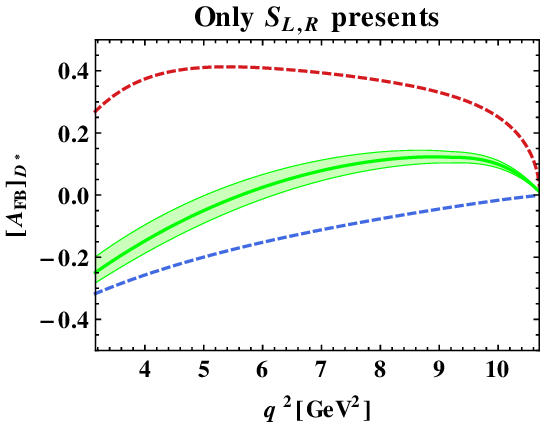}\,\includegraphics[width=4.5cm]{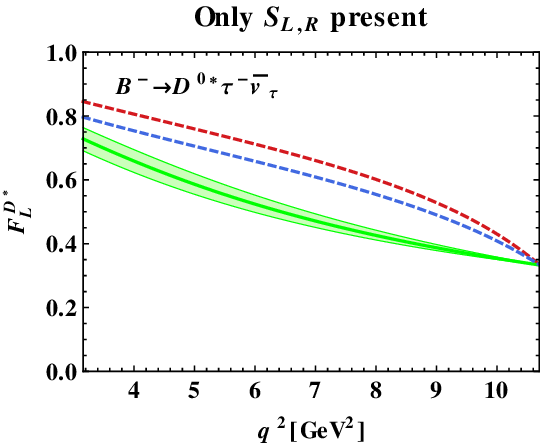}\,
~~\includegraphics[width=4.5cm]{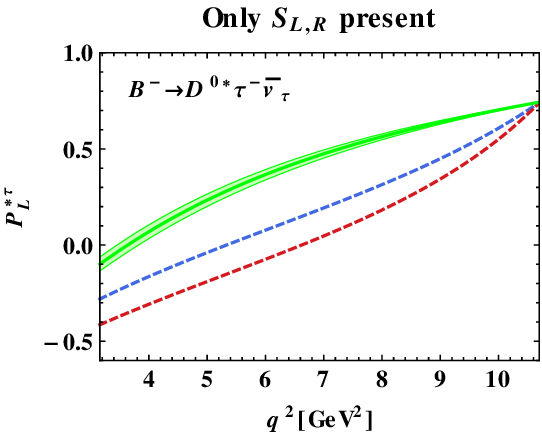}
\caption{The left (center) and right panel shows the $q^2$ dependence of $[A_{FB}]_{D^*}$ ($F^{D^*}_L$) and $P^{* \tau}_L$  for the decay $\bar{B} \to D^{0*} \tau^- \bar{\nu}_\tau$.  The dashed lines show predictions for some  representative values of the new couplings $S_L$ and $S_R$. For example the red lines corresponds to $(S_L, S_R) = (-1.93, 1.73)$ in the left panel,  $(S_L, S_R) = (0.80, -0.9)$ in the middle panel and $(S_L, S_R) = (0.61, -0.98)$ in the right panel.
\label{fig:AFBDstonlygP}}
\end{figure}

The transverse forward-backward asymmetry $[A^{T}_{FB}]_{D^*}$  is not sensitive to the $S_L$ and $S_R$  couplings . In  Fig.~\ref{fig:AFBDstonlygP} we show the effects of $S_L$ and $S_R$  on $[A_{FB}]_{D^{*} }$, the polarization fractions $F^{D^*}_L(q^2)$ and $P^{*\tau}_L(q^2)$.  These couplings  can positively or negatively enhance $[A_{FB}]_{D^{*} }$ to about $30\%$ at low $q^2$ and there can be different zero-crossing than  the SM.  Unlike the $V_L$ and $V_R$  case, the polarization fractions are sensitive to the $S_L$ and $S_R$ couplings. Due to the  $S_L$ and $S_R$ couplings, $F^{D^*}_L(q^2)$ can be as large as 0.85 at low $q^2$, and it decreases to the SM value at high $q^2$. The polarization fraction $P^{* \tau}_L(q^2)$ can be as large as 0.5 and negative at very low $q^2$ and can have different zero-crossing than the SM. It increases to the SM value at high $q^2$.
 
\subsection{$\bar{B} \to D^{0} \tau^- \bar{\nu}_\tau$}
\subsubsection{Only $g_V$ coupling present}
We now consider predictions for the various observables in
$\bar{B} \to D^{0} \tau^- \bar{\nu}_\tau$. The axial vector coupling $g_A$ does not contribute in this case and hence we consider only the  coupling $g_V$.
Note that, the forward-backward asymmetry and the $\tau$ polarization fraction are independent of the  coupling $g_V$.

\begin{figure}[h!]
\centering
\includegraphics[width=5.5cm]{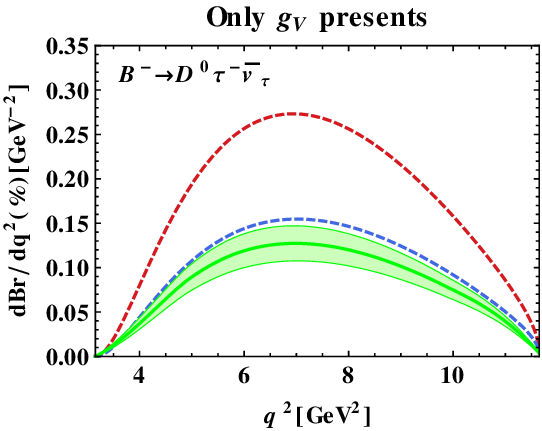}\,~~\includegraphics[width=5.5cm]{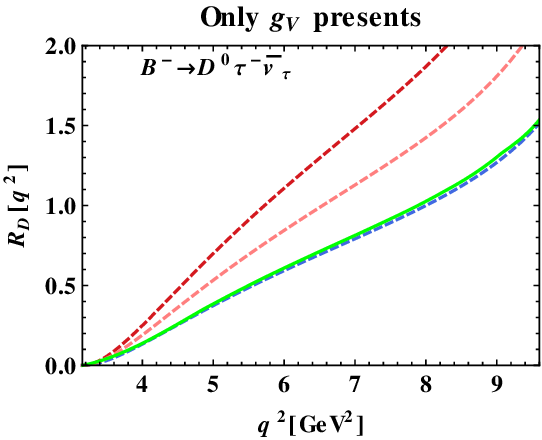}
\caption{The left (right) panel shows the $q^2$ dependence of DBR ($R_{D}(q^2)$)  for the decay $\bar{B} \to D^0 \tau^- \bar{\nu}_\tau$. The dashed lines show predictions for some  representative values of $g_V$. For example the red lines  correspond to $g _V = 0.68 e^{i 1.27}$ in the left panel and $g _V = 0.62 e^{i 1.20}$ in the right panel.  \label{fig:DdBrRDonlygV}}
\end{figure}

In the presence of $g_V$ ,  the $q^2$ dependence of DBR and $R_{D}(q^2)$ for  the   decay $\bar{B} \to D^0 \tau^- \bar{\nu}_\tau$  are shown in Fig.~\ref{fig:DdBrRDonlygV}.
The  DBR can increase upto 0.25\% at $q^2 \approx 7 \mathrm{GeV^2}$.  The ratio $R_{D}(q^2)$ is proportional to $(1 + g_V)^2$ and increases with  $q^2$.

\begin{figure}[h!]
\centering
\includegraphics[width=5.5cm]{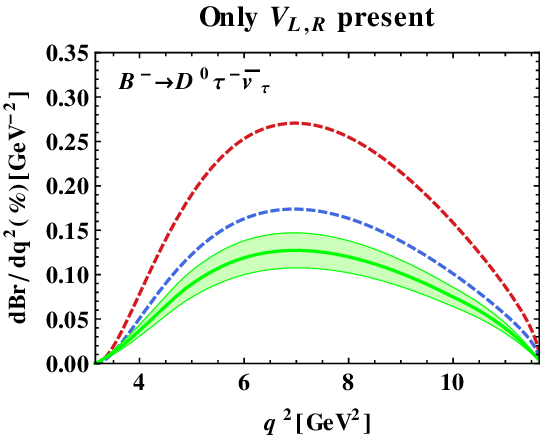}\,
~~\includegraphics[width=5.5cm]{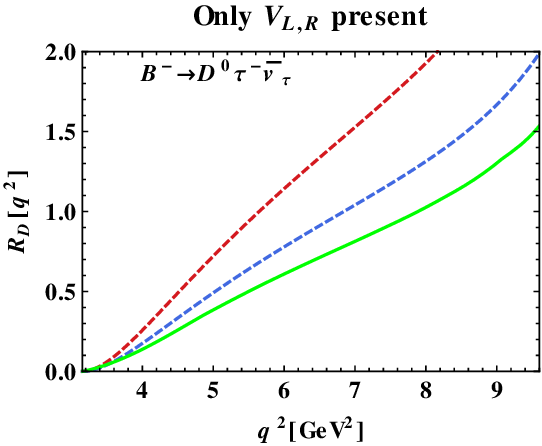}
\caption{The left (right) panel shows the $q^2$ dependence of DBR ($R_{D}(q^2)$)  for the decay $\bar{B} \to D^0 \tau^- \bar{\nu}_\tau$. The dashed lines show predictions for some  representative values of new couplings $V_L $ and $ V_R$. For example the red lines correspond to $(V_L , V_R) = (-1.09, -1.28)$. \label{fig:DdBrRDonlyLR}}
\end{figure}

Now we consider $V_{L,R}$ real and independent.  The $q^2$ dependence of DBR and $R_{D}(q^2)$ for  the   decay $\bar{B} \to D^0 \tau^- \bar{\nu}_\tau$  are shown in Fig.~\ref{fig:DdBrRDonlyLR}. The predicted deviations from the SM are similar to the case with pure $g_V$ coupling.

\subsubsection{Pure $S_L$ and $S_R$ couplings present}
Finally, we consider the effect of $S_{L,R}$ , taken to be real, on the observables in 
$\bar{B} \to D^0 \tau^- \bar{\nu}_\tau$. The allowed ranges for the real $S_L$ and $S_R$ couplings are shown as the colored region of Fig.~\ref{fig:OnlygSgP}(right panel).

\begin{figure}[h!]
\centering
\includegraphics[width=5.5cm]{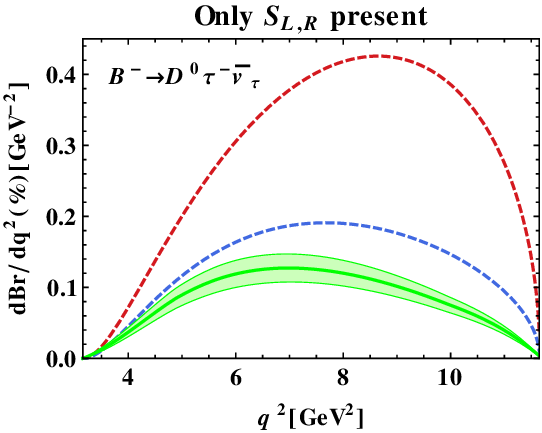}\,
~~\includegraphics[width=5.5cm]{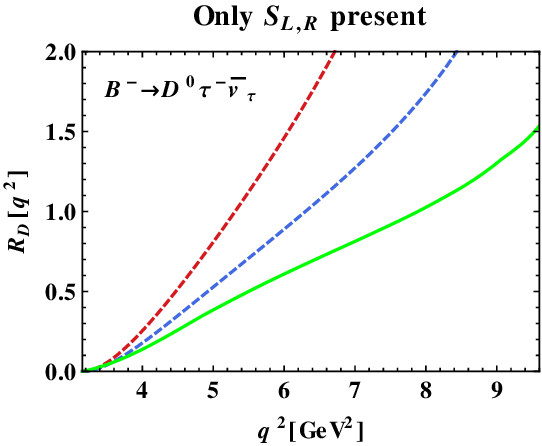}
\caption{The left (right) panels of the figure show the $q^2$ dependence of DBR ($R_{D}(q^2)$)   for the decay $\bar{B} \to D^0 \tau^- \bar{\nu}_\tau$. The dashed lines show predictions for some  representative values of new couplings $S_L$ and $S_R$. For example the red lines  correspond to $(S_L, S_R) = (0.51, -1.29)$.
\label{fig:DdBrRDonlygSgP}}
\end{figure}

In Fig.~\ref{fig:DdBrRDonlygSgP} we show the effect of $S_L$ and $S_R$ couplings on the DBR and $R_{D}(q^2)$ in the decay $\bar{B} \to D^0 \tau^- \bar{\nu}_\tau$. The DBR increases up to 0.25\%$\rm GeV^{-2}$ at $q^2 \approx 9.5 \mathrm{GeV^2}$. Note that the peak of the distribution in the DBR can be shifted towards high $q^2$ relative to the SM. The deviation in the ratio $R_{D}(q^2)$ increases with $q^2$.

\begin{figure}[h!]
\centering
\includegraphics[width=5.5cm]{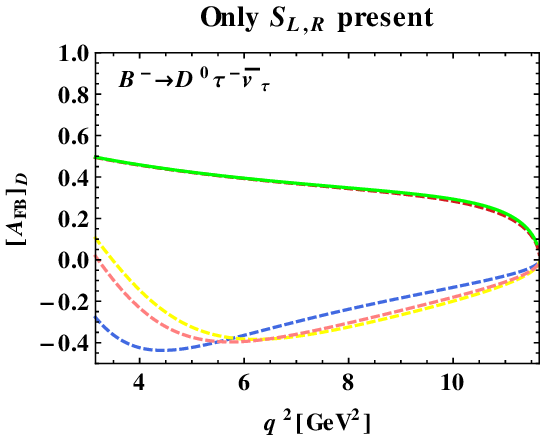}\,
~~\includegraphics[width=5.5cm]{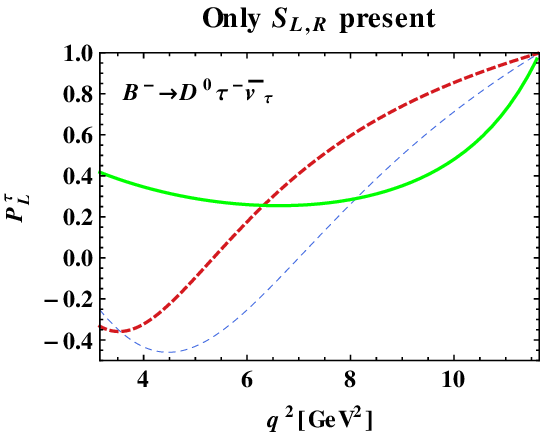}
\caption{The left (right) panel shows the $q^2$ dependence of $[A_{FB}]_{D}$ ($P^{ \tau}_L$ ) for the decay $\bar{B} \to D^{0} \tau^- \bar{\nu}_\tau$.  The dashed lines show predictions for some  representative values of new couplings $S_L$ and $S_R$. For example the red lines correspond to $(S_L, S_R) = (2.11, -0.04)$ in the left panel and $(S_L, S_R) = (1.42, 0.21)$ in the right panel.
\label{fig:AFBDonlygS}}
\end{figure}

Unlike the $V_{L,R}$ case, the forward-backward asymmetry and $\tau$-polarization fraction in the decay $\bar{B} \to D^{0} \tau^- \bar{\nu}_\tau$ are very sensitive to the $S_L$ and $S_R$ couplings as shown in Fig.~\ref{fig:AFBDonlygS}.  In this case,  $[A_{FB}]_{D}$ can be either positive or negative, and may have zero-crossing. In the SM , there is no zero-crossing for  $[A_{FB}]_{D}$ . The $\tau$-polarization fraction can be negatively enhanced to more than 40\% at low $q^2$ and may have the zero-crossing.

\section{Summary}
In summary, we have considered new physics explanation of the recent measurements in
$\bar{B} \to D^{0*} \tau^- \bar{\nu}_\tau$ and $\bar{B} \to D \tau^- \bar{\nu}_\tau$  
decays by the BaBar collaboration. We considered an effective Lagrangian description of the new physics with four-fermi operators with vector/axial vector and scalar/pseudoscalar couplings. We considered two cases, in the first case we considered only V/A couplings and in the second case we considered only S/P couplings. We found that the measurements in the decay $\bar{B} \to D^{0*} \tau^- \bar{\nu}_\tau$ strongly constrain the new physics couplings to be only pure vector or pure axial vector. 
Assuming a pure vector and pure axial vector complex couplings we calculated the differential branching ratios, the ratio $R_{D^{*}}(q^2)$, the forward-backward  asymmetries and  the polarization fractions of the tau and the $D^*$ meson. We found that the pure complex vector couplings $g_V$ only affects the forward-backward asymmetries. The complex axial vector couplings, on the other hand, affects the DBR, $R_{D^{*}}(q^2)$ as well as the
forward-backward asymmetries. The polarization fractions are not affected in presence of the $g_{V,A}$ couplings. When we considered real scalar couplings $S_{L,R}$ , we found that all observables except the transverse forward-backward asymmetry for the $D^*$, $[A^{T}_{FB}]_{D^*}$ were affected.  Moving on to the decay $\bar{B} \to D^0 \tau^- \bar{\nu}_\tau$ we found that the axial vector new physics coupling do not contribute to this decay. In the case of the vector coupling $g_V$, the forward-backward asymmetry and the tau polarization fraction were not affected though the DBR and the ratio $R_D(q^2)$ were affected. In the presence of the scalar couplings $S_{L,R}$, not only the DBR and the ratio $R_D(q^2)$ were affected but the forward-backward asymmetry and the tau  polarization fraction were also affected and were found to be very sensitive to the scalar couplings. The fact that different new physics couplings have different effects on the observables  demonstrated that by measuring the various observables it is possible to distinguish different models of new physics.
Finally, if new physics is established in both $R(D)$ and $ R(D^*)$ then the cases of pure $g_A= V_R-V_L$  and pure $S_R-S_L$ couplings are ruled out as they contribute to only $ R(D^*)$  and the pure $S_R+S_L$ is ruled out as it contributes to only $R(D)$ .

\section*{Acknowledgements} We thank Gilad Perez, M. Papucci, M.~Gonzalez-Alonso
 and J.~F.~Kamenik for useful discussion and comments. DG  thanks Amol Dighe for discussion and encouragement. This work was financially supported  by the US-Egypt Joint Board on Scientific and
Technological Co-operation award (Project ID: 1855) administered by
the US Department of Agriculture and in part by the National Science
Foundation under Grant No.\ NSF PHY-1068052. AD thanks the hospitality of the CERN theory group where the work was completed.

\appendix
\section*{\centering Appendix}

\section{Kinematics}
\label{kinematics}
The matrix element square for the decay $\bar{B} \to D^{(*)} l \nu_l$ can be factorized into leptonic ($L_{\mu \nu}$) and hadronic ($H_{\mu \nu}$) tensors as
\bea
\label{eq3:ME}
|{\cal{M}}(\bar{B} \to D^{(*)}  l \nu_l)|^2 &=& |\bra{D^{(*)}  l \nu}{\cal{L}}_{eff} \ket{\bar{B}}|^2 = L_{\mu \nu} H^{\mu \nu} \,.
 \eea
The polar angle $\cos{\theta_l}$ dependence of the leptonic  and hadronic  tensors  $ L_{\mu \nu} H^{\mu \nu}$   can be evaluated using the completeness relation for the polarization four-vectors $\bar{\epsilon}(m = 0,\pm,t)$ \cite{Korner:1989qb,Kadeer:2005aq}:
\bea
\label{eq4:cpol}
\sum_{m, m^\prime  = 0,\pm,t} \bar{\epsilon}^\mu(m) \bar{\epsilon}^{*\nu }(m^\prime) g_{m m^\prime}&=& g^{\mu\nu} \,,
\eea
where the tensor $g_{m m^\prime} = diag(+,-,-,-)$. The matrix element square reduces to
\bea
\label{eq4:ME}
|{\cal{M}}(\bar{B} \to D^{(*)}  l \nu_l)|^2 &=& \sum_{m,m^\prime,n,n^\prime} L(m,n) H(m^\prime,n^\prime)  g_{mm^\prime} g_{nn^\prime} \,,
 \eea
where $L(m,n) = L^{\mu \nu } \bar{\epsilon}_\mu(m) \bar{\epsilon}^*_\nu(n) $ and $H(m,n) = H^{\mu \nu } \bar{\epsilon}^*_\mu(m) \bar{\epsilon}_\nu(n) $. The advantage of Eq.\ref{eq4:ME} is that
$L(m,n)$ and  $H(m^\prime,n^\prime)$ are
Lorentz invariant and so one can evaluate them in different Lorentz frames \cite{Kadeer:2005aq}. The leptonic tensor $L[m,n]$  will be evaluated in the $l-\nu_l$ center-of-mass (c.m.) frame ($q^2$ rest frame), whereas the hadroic tensor $H[m,n]$ in the B rest frame.

In $B$ rest frame, we choose  the helicity basis $\bar{\epsilon}$
\bea
\label{eqAppK1:hlvecB}
\bar{\epsilon}(0) &=& \frac{1}{\sqrt{q^2}}(|p_{D^{(*)}}|,0,0,-q_0)\,,\quad 
\bar{\epsilon}(\pm) = \pm \frac{1}{\sqrt{2}}(0,\pm 1, -i ,0)\,,\nl
\bar{\epsilon}(t) &=& \frac{1}{\sqrt{q^2}}(q_0,0,0,|p_{D^{(*)}}|)\,,
\eea
where $q_0 = (m^2_B - m^2_{D^*} + q^2)/2 m_B$ and $|p_{D^{(*)}}| = \lambda^{1/2}(m^2_B,m^2_{D^{(*)}},q^2)/2 m_B$. In this frame,   the B and  $D^{(*)}$ mesons four-momenta  $p_B$ and $p_{D^{(*)}}$ are
\bea
\label{eqAppK2:MomB}
p_B &=& (m_B,0,0,0)\,,~~p_{D^{(*)}} = (E_{D^{(*)}},0,0,|p_{D^{(*)}}|)\,,
\eea
where $E_{D^{(*)}} = (m^2_B + m^2_{D^{(*)}} -q^2)/ 2 m_B$. The momentum transfer  q is $q=p_B-p_{D^{(*)}}$.  Further, one chooses the polarization vector of the $D^*$ meson as
\bea
\label{eqAppK3:hlvecDstB}
\bar{\epsilon}(0) &=& \frac{1}{m_{D^*}}(|p_{D^*}|,0,0, E_{D^*})\,,\quad
\bar{\epsilon}(\pm)= \mp \frac{1}{\sqrt{2}}(0, 1, \pm i ,0)\,.
\eea

The leptonic tensor $L[m,n]$ will be evaluated in the $q^2$ rest frame. In this frame, we choose  the  transverse components of helicity basis $\bar{\epsilon}$  to remain the same and other two components are taken as
\bea
\label{eqAppK4:hlvecLep}
\bar{\epsilon}(0) &=& (0,0,0,-1)\,,~
\bar{\epsilon}(t)=(1,0,0,0)\,.
\eea

Let $\theta_l$ be the angle between the $D^{(*)}$ meson and the $\tau$ lepton three-momenta in the $q^2$ rest frame. We define the momenta of the lepton and anti-neutrino  pair as
\bea
\label{eqAppK5:momLep}
p^\mu_l &=& (E_l, pl \sin{\theta_l}, 0, -pl \cos{\theta_l})\,,\nl
p^\mu_\nu &=& (p_l, -pl \sin{\theta_l}, 0, pl \cos{\theta_l})\,,
\eea
where the lepton energy $E_l = (q^2 + m^2_l)/2 \sqrt{q^2}$ and magnitude of its three-momenta  is $p_l =  (q^2 - m^2_l)/2 \sqrt{q^2}$.

\section{$\BDstartaunu$ details}
\label{Dstardetails}
The $\BDstartaunu$  differential decay rates for the lepton helicity $\lambda_\tau = \pm \frac{1}{2}$   are

\bea
\label{eqApp1:DDRBstdtn}
\frac{d \Gamma^{D^*}[\lambda_\tau = -1/2]}{dq^2 d\cos{\theta_l}}&=& N |p_{D^*}| \Big[ 2 |{\cal{A}}_0|^2  \sin^2{\theta_l} + (|{\cal{A}}_\parallel|^2+ |{\cal{A}}_\perp|^2) (1 + \cos{\theta_l}^2) -4 Re[{\cal{A}}_\parallel {\cal{A}}^*_\perp] \cos{\theta_l} \Big]\,,\nl
\frac{d \Gamma^{D^*}[\lambda_\tau = 1/2]}{dq^2 d\cos{\theta_l}}&=&  N |p_{D^*}| \frac{m^2_\tau}{q^2}  \Big[  2 |{\cal{A}}_0 \cos{\theta_l}-{\cal{A}}_{tP}|^2  + (|{\cal{A}}_\parallel|^2+ |{\cal{A}}_\perp|^2)  \sin^2{\theta_l} \Big]\,.
\eea
The differential decay rate corresponding to the helicity $\lambda_\tau = 1/2$  vanishes for the light leptons $(e,\mu)$.

The relevant form factors for the $B \to D^*$ matrix elements of the vector $V_\mu = \bar{c}\gamma^{\mu}b$ and  axial-vector  $A_\mu = \bar{c}\gamma^{\mu} \gamma_5 b$ currents are defined as \cite{Beneke:2000wa}
\bea
\label{eqApp3:MEVFF}
  \bra{ D^*(p_{D^*},\epsilon^*)}V_\mu \ket{\bar{B}(p_B)}  &=&
 \frac{2 i V(q^2)}{m_B + m_{D^*}}\varepsilon_{\mu \nu \rho \sigma} \epsilon^{*\nu}  p^{\rho}_{D^*} p^{\sigma}_B \,,\nl 
 \bra{ D^*(p_{D^*},\epsilon^*)}A_\mu \ket{\bar{B}(p_B)}  &=&  2 m_{D^*} A_0 (q^2)\frac{\epsilon^*.q}{q^2} q_\mu + (m_B + m_{D^*}) A_1(q^2) \Big[\epsilon^*_{\mu}-\frac{\epsilon^*.q}{q^2} q_\mu \Big]\nl && \hspace*{-4.5cm}-A_2(q^2) \frac{\epsilon^*.q}{(m_B + m_{D^*})} \Big[(p_B +p_{D^*})_\mu -\frac{m^2_B-m^2_{D^*}}{q^2}q_\mu \Big]\,.
\eea

In addition,  from  Eq.~(\ref{eqApp3:MEVFF}) one can show that  the $B \to D^*$ matrix element for the scalar current vanishes and for the pseudoscalar current reduces to

\bea
\label{eqApp3:MEVFF2}
\bra{ D^*(p_{D^*},\epsilon^*)}\bar{c}\gamma_5 b\ket{\bar{B}(p_B)}  &=& -\frac{ 2 m_{D^*} A_0 (q^2)}{m_b(\mu) + m_c(\mu)}\epsilon^*.q\,.
\eea

The expression of the  hadronic helicity amplitudes for the $\barBstdtn$ decays are
\bea
\label{eqApp1:BDstAmp}
{\cal{A}}_0  &=&\frac{1}{2 m_{D^*} \sqrt{q^2}} \Big[(m_B^2 - m_{D^*}^2 - q^2) (m_B + m_{D^*} ) A_1(q^2) -\frac{4 m_B^2 |p_{D^*}|^2}{m_B +m_{D^*}}  A_2(q^2) \Big](1 - g_A) \,,\nl
{\cal{A}}_{\pm}  &=& \Big[(m_B + m_{D^*})A_1(q^2) (1 - g_A)  \mp \frac{2 m_B V(q^2)}{(m_B + m_{D^*})} |p_{D^*}|(1 + g_V) \Big]\,,\nl
{\cal{A}}_{t}  &=& \frac{2 m_B |p_{D^*}| A_0(q^2) }{\sqrt{q^2}} (1 - g_A) \,,\nl
{\cal{A}}_{P}  &=& -\frac{2 m_B |p_{D^*}| A_0(q^2) }{ (m_b(\mu) + m_c(\mu))} g_P \,.
\eea
\section{$\BDtaunu$ results}
\label{Ddetails}
The $\barBdtn$  differential decay rates for the lepton helicity $\lambda_\tau = \pm \frac{1}{2}$   are

\bea
\label{eqApp1:DDRBdtn}
\frac{d \Gamma^D[\lambda_\tau = -1/2]}{dq^2 d\cos{\theta_l}}&=& 2 N |p_D|  |H_0|^2  \sin^2{\theta_l} \,,\nl
\frac{d \Gamma^D[\lambda_\tau = 1/2]}{dq^2 d\cos{\theta_l}}&=& 2 N |p_D|  \frac{m^2_\tau}{q^2} |H_0 \cos{\theta_l}-H_{tS}|^2 \,.
\eea
The differential decay rate corresponding to the helicity $\lambda_\tau = 1/2$  vanishes for the light leptons $(e,\mu)$.

The pseudoscalar form factors $F_+(q^2)$ and $F_0(q^2)$ of the  $B \to D$ matrix elements are defined as
\bea
\label{eqApp2:MEBD}
  \bra{ D(p_D)}\bar{c}\gamma^{\mu}b\ket{\bar{B}(p_B)}  &=&
  F_+(q^2)~ 
  \Big[p_B^{\mu}+p_D^{\mu}-\frac{m_B^2-m^2_D}{q^2}q^{\mu}\Big] 
  +  F_0(q^2)~\frac{m_B^2-m^2_D}{q^2}q^{\mu}\,,\nl 
 \bra{ D(p_D)}\bar{c}b\ket{\bar{B}(p_B)}  &=&  \frac{m_B^2-m^2_D}{
   m_b (\mu) - m_c(\mu)}F_0(q^2)\,.
\eea

The helicity amplitudes are

\bea
\label{eqApp3:BDAmp}
H_0 &=& \frac{2 m_B |p_D|}{\sqrt{q^2}} F_+(q^2) (1 + g_V )\,,\quad
H_t = \frac{m^2_B -m^2_D}{\sqrt{q^2}}  F_0(q^2)   (1 + g_V )\,,\nl
H_S &=& \frac{m^2_B -m^2_D}{m_b (\mu) - m_c(\mu)}  F_0(q^2)  g_S\,.
\eea

\section{ Form factors in the Heavy Quark Effective Theory}
\label{FF}
In the heavy quark limit for the b, c quarks $(m_{b,c} \gg \Lambda_{QCD})$, both charm
and the bottom quark in the hadronic current have to be replaced by static quarks  $h_{v^\prime,c}$ and $h_{v,b}$, where $v^\mu_B = p_B/m_B$ and $v^{\prime \mu}_{D^{(*)}} = p_{D^{(*)}}/m_{D^{(*)}} $ are the four-velocity of the B and $D(D^*)$ mesons, respectively. The $ b \to c $ transition can be studied in the heavy quark effective theory (HQET). 
In this effective theory, the matrix elements of the vector and axial vector currents, $V_\mu$ and 
and $A_\mu$ , between bottom and charm mesons \cite{Falk:1992wt} are defined as
\bea
\label{eqApp5:MEVFFHL}
  \langle D(v') |\,V_\mu\,| B(v) \rangle &=&
    \sqrt{m_B m_D}\,\Big[ h_+(w)\,(v+v')_\mu  +
    h_-(w)\,(v-v')_\mu \Big] \,, \nonumber\\
\langle D^*(v',\epsilon') |\,V_\mu\,| B(v) \rangle &=&
    i \sqrt{m_B m_{D^*}}\,\,h_V(w)\,
    \epsilon_{\mu\nu\alpha\beta}
    \,\epsilon'^{*\nu}\,v'^\alpha\,v^\beta \,, \nonumber\\
   && \nonumber\\
   \langle D^*(v',\epsilon') |\,A_\mu\,| B(v) \rangle &=&
    \sqrt{m_B m_{D^*}}\,\Big[ h_{A_1}(w)\,(w+1)\,
    \epsilon_\mu'^*  - h_{A_2}(w)\,\epsilon'^*\!\!\cdot\! v\,v_\mu
   \nonumber\\
   && - h_{A_3}(w)\,\epsilon'^*\!\!\cdot\! v\,v'_\mu \Big] \,,
\eea
where the  kinematical variable $w =v_B.v_{D^{(*)}} = (m^2_B + m^2_{D^{(*)}}-q^2)/2 m_B m_{D^{(*)}}$. \vspace*{1mm}

The form factors $ F_+(q^2)$ and $F_0(q^2)$ in Eq.~(\ref{eqApp2:MEBD}) are related to the form factors $h_+(w)$ and $h_-(w)$ via
\bea
\label{eq:BDFFrel}
F_+(q^2) &=& \frac{V_1(w)}{R_D}\,,\quad
F_0(q^2)= \frac{(1 + w) R_D}{2}  S_1(w)\,,
\eea
where 
\bea
\label{eq:V1S1exp}
V_1(w)  &=& \Big[h_+(w) - \frac{(1 - r)}{(1 + r)} h_-(w)\Big]\,,\nl
 S_1(w)  &=& \Big[h_+(w) -  \frac{(1 + r)}{(1 - r)} \frac{(w-1)}{(w+1)}   h_-(w)\Big]\,.
\eea
Here $R_{D} = 2 \sqrt{m_B m_D}/(m_B + m_D)$ and $r = m_D/m_B$. We will use the parametrization of the  form factor $V_1(w)$  as given by \cite{Caprini:1997mu}
\bea
\label{eq:V1exp}
V_1(w)  &=& V_1(1) [1 -8 \rho^2_1 z + (51 \rho^2_1 - 10)z^2  - (252 \rho^2_1- 84) z^3 ]\,,
\eea
where $z =( \sqrt{w+1}-\sqrt{2})/( \sqrt{w+1}+\sqrt{2})$. The numerical values of the free parameters are \cite{Aubert:2009ac}
\bea
\label{eq:V1thpsqnum}
V_1(1) |V_{cb}| &=& (43.0 \pm 1.9 \pm 1.4)\times 10^{-3}\,,\nl
\rho^2_1 &=& 1.20 \pm 0.09 \pm 0.04.
\eea

For the form factor $ S_1(w)$ we employ the parameterization as in \cite{Sakaki:2012ft},
\bea
\label{eq:S1exp}
S_1(w)  &=&  [1.0036 - 0.0068(w − 1) + 0.0017(w − 1)^2] V_1(w) \,.
\eea

In the HQET, the helicity amplitudes in 
Eq.~(\ref{eqApp1:DDRBdtn}) becomes
\bea
\label{eq:BDlnuHLAmp}
H_0 &=& m_B (1 + r) \sqrt{\frac{r (w^2 - 1)}{(1 + r^2 - 2 r w)}} V_1[w](1 + g_V)\,,\nl
H_{tS} &=& \frac{m_B (1 - r) \sqrt{r} (w + 1)}{\sqrt{(1 + r^2 - 2 r w)}}    S_1[w]  \Big[(1 + g_V)-\frac{m_B^2 (1 + r^2 - 2 r w) }{m_l(m_b(\mu) - m_c(\mu)) } g_S \Big]\,.
\eea


\vspace*{1mm}
Now we consider the form factors for the $B \to D^*$ matrix element in the HQET. The form factors $h_{A_i}(w)$ are related to the form factors in Eq.~(\ref{eqApp3:MEVFF})  \cite{ Fajfer:2012vx, Caprini:1997mu,Dungel:2010uk} in the following way,
\bea
\label{eqApp6:MEVFFHL2}
A_1(q^2) &= & R_{D^*} \frac{w+1}{2}h_{A_1}(w)\,,\quad
A_0(q^2) = \frac{R_0(w)}{R_{D^*}}h_{A_1}(w)\,,\nl
A_2(q^2) &= & \frac{R_2(w)}{R_{D^*}}h_{A_1}(w)\,,\quad
V(q^2)= \frac{R_1(w)}{R_{D^*}}h_{A_1}(w)\,,
\eea
where $R_{D^*} = 2 \sqrt{m_B m_D^*}/(m_B + m_D^*)$. The $w$ dependence of the form factors can be found in \cite{Fajfer:2012vx, Caprini:1997mu} and the summary of the results are
\bea
\label{eqApp7:MEVFFHL3}
h_{A_1}(w) &=& h_{A_1}(1)\Big[1-8 \rho^2 z + (53 \rho^2-15)z^2  -(231 \rho^2 -91) z^3\Big]\,,\nl
R_1(w)  &=& R_1(1) - 0.12 (w-1) + 0.05 (w-1)^2 \,,\nl
R_2(w)  &=& R_2(1) + 0.11 (w-1) - 0.06(w-1)^2 \,,\nl
R_0(w)  &=& R_0(1) - 0.11 (w-1) + 0.01(w-1)^2 \,,
\eea
where $z =( \sqrt{w+1}-\sqrt{2})/( \sqrt{w+1}+\sqrt{2})$. The numerical values of the free parameters $\rho^2$, $h_{A_1}(1)$, $R_1(1)$ and $R_2(1) $ are taken from\cite{Dungel:2010uk},
\bea
\label{eqApp8:MEVFFHLNum}
h_{A_1}(1) |V_{cb}| &=& (34.6\pm  0.2 \pm 1.0) \times 10^{-3}\,,\nl
\rho^2 &=& 1.214 \pm 0.034 \pm 0.009\,,\nl
R_1(1) &=& 1.401\pm 0.034 \pm 0.018\,,\nl
R_2(1) &=& 0.864 \pm 0.024 \pm 0.008 \,,
\eea
and $R_0(1) = 1.14$ is taken from\cite{Fajfer:2012vx}. In the numerical analysis, we allow  $10 \%$ uncertainties in the $R_0(1)$ value to account  higher order corrections.

In the HQET, the transversity amplitudes of Eq.~(\ref{eqApp1:BDstAmp}) become
\bea
\label{eq:BDstAmpHL}
{\cal{A}}_0  &=& \frac{ m_B  (1 - r_*) (w + 1) \sqrt{r_*}}{ \sqrt{ (1 + r^2_* - 2 r_* w)}}h_{A_1}(w) \Big[1 + \frac{(w-1)(1-R_2(w))}{(1 - r_*)}\Big](1-g_A)\,,\nl
{\cal{A}}_\| &=&  m_B \sqrt{2r_*}(w + 1)h_{A_1}(w)(1 - g_A)\,,\nl
{\cal{A}}_\perp &=& - m_B   \sqrt{2r_* (w^2 - 1)} h_{A_1}(w)R_1(w) (1 + g_V)\,,
\eea
where $r_* = m_{D^*}/m_B$.


\begin{thebibliography}{00}
\bibitem{belletau}
K.~Ikado {\it et al.},
  Phys.\ Rev.\ Lett.\  {\bf 97}, 251802 (2006)
  [arXiv:hep-ex/0604018].
\bibitem{Bhattacherjee:2010ju} 
See for example, 
B.~Bhattacherjee, A.~Dighe, D.~Ghosh and S.~Raychaudhuri,
  Phys.\ Rev.\ D {\bf 83}, 094026 (2011)
  [arXiv:1012.1052 [hep-ph]].

\bibitem{MartinGon}
A.~Filipuzzi, J.~Portoles and M.~Gonzalez-Alonso,
  arXiv:1203.2092 [hep-ph].

\bibitem{nierste} 
  U.~Nierste, S.~Trine and S.~Westhoff,
  Phys.\ Rev.\ D {\bf 78}, 015006 (2008)
  [arXiv:0801.4938 [hep-ph]].
\bibitem{Matyja:2007kt} 
  A.~Matyja {\it et al.}  [Belle Collaboration],
  Phys.\ Rev.\ Lett.\  {\bf 99}, 191807 (2007)
  [arXiv:0706.4429 [hep-ex]].
\bibitem{Aubert:2007dsa} 
  B.~Aubert {\it et al.}  [BABAR Collaboration],
  Phys.\ Rev.\ Lett.\  {\bf 100}, 021801 (2008)
  [arXiv:0709.1698 [hep-ex]].
\bibitem{Adachi:2009qg} 
  I.~Adachi {\it et al.}  [Belle Collaboration],
  arXiv:0910.4301 [hep-ex].
\bibitem{Bozek:2010xy} 
  A.~Bozek {\it et al.}  [Belle Collaboration],
  Phys.\ Rev.\ D {\bf 82}, 072005 (2010)
  [arXiv:1005.2302 [hep-ex]].
\bibitem{:2012xj} 
  [BaBar Collaboration],
  arXiv:1205.5442 [hep-ex].
\bibitem{Fajfer:2012vx} 
  S.~Fajfer, J.~F.~Kamenik and I.~Nisandzic,
  arXiv:1203.2654 [hep-ph].
\bibitem{Sakaki:2012ft} 
  Y.~Sakaki and H.~Tanaka,
  arXiv:1205.4908 [hep-ph].
\bibitem{npbabar}
S.~Fajfer, J.~F.~Kamenik, I.~Nisandzic and J.~Zupan,
  arXiv:1206.1872 [hep-ph];
A.~Crivellin, C.~Greub and A.~Kokulu,
  arXiv:1206.2634 [hep-ph].
\bibitem{ccLag} 
  T.~Bhattacharya, V.~Cirigliano, S.~D.~Cohen, A.~Filipuzzi, M.~Gonzalez-Alonso, M.~L.~Graesser, R.~Gupta and H.~-W.~Lin,
  Phys.\ Rev.\ D {\bf 85}, 054512 (2012)
  [arXiv:1110.6448 [hep-ph]];
  C.~-H.~Chen and C.~-Q.~Geng,
  Phys.\ Rev.\ D {\bf 71}, 077501 (2005)
  [hep-ph/0503123].
\bibitem{dattaneutrino} 
See for example,
A.~Rashed, M.~Duraisamy and A.~Datta,
  arXiv:1204.2023 [hep-ph];
A.~Datta, P.~J.~O'Donnell, Z.~H.~Lin, X.~Zhang and T.~Huang,
  Phys.\ Lett.\ B {\bf 483}, 203 (2000)
  [hep-ph/0001059].
B.~Bhattacherjee, S.~S.~Biswal and D.~Ghosh,
  Phys.\ Rev.\ D {\bf 83}, 091501 (2011)
  [arXiv:1102.0545 [hep-ph]].
\bibitem{dattaBnp} See for e.g.
A.~Datta,
  Phys.\ Rev.\ D {\bf 66}, 071702 (2002)
  [hep-ph/0208016];
A.~Datta and P.~J.~O'Donnell,
  Phys.\ Rev.\ D {\bf 72}, 113002 (2005)
  [hep-ph/0508314];
A.~Datta,
  Phys.\ Rev.\ D {\bf 74}, 014022 (2006)
  [hep-ph/0605039];
A.~Datta,
  Phys.\ Rev.\ D {\bf 78}, 095004 (2008)
  [arXiv:0807.0795 [hep-ph]];
C.~-W.~Chiang, A.~Datta, M.~Duraisamy, D.~London, M.~Nagashima and A.~Szynkman,
  JHEP {\bf 1004}, 031 (2010)
  [arXiv:0910.2929 [hep-ph]].
\bibitem{Neubert:1993mb} 
  M.~Neubert,
  Phys.\ Rept.\  {\bf 245}, 259 (1994)
  [hep-ph/9306320].
\bibitem{Caprini:1997mu} 
  I.~Caprini, L.~Lellouch and M.~Neubert,
  Nucl.\ Phys.\ B {\bf 530}, 153 (1998)
  [hep-ph/9712417].
\bibitem{Nakamura:2010zzi} 
  K.~Nakamura {\it et al.}  [Particle Data Group Collaboration],
  J.\ Phys.\ G G {\bf 37}, 075021 (2010).
\bibitem{Asner:2010qj} 
  D.~Asner {\it et al.}  [Heavy Flavor Averaging Group Collaboration],
  arXiv:1010.1589 [hep-ex].
\bibitem{Korner:1989qb} 
  J.~G.~Korner and G.~A.~Schuler,
  Z.\ Phys.\ C {\bf 46}, 93 (1990).
\bibitem{Kadeer:2005aq} 
  A.~Kadeer, J.~G.~Korner and U.~Moosbrugger,
  Eur.\ Phys.\ J.\ C {\bf 59}, 27 (2009)
  [hep-ph/0511019].
\bibitem{Beneke:2000wa} 
  M.~Beneke and T.~Feldmann,
  Nucl.\ Phys.\ B {\bf 592}, 3 (2001)
  [hep-ph/0008255].
\bibitem{Falk:1992wt} 
  A.~F.~Falk and M.~Neubert,
  Phys.\ Rev.\ D {\bf 47}, 2965 (1993)
  [hep-ph/9209268].
\bibitem{Aubert:2009ac} 
  B.~Aubert {\it et al.}  [BABAR Collaboration],
  Phys.\ Rev.\ Lett.\  {\bf 104}, 011802 (2010)
  [arXiv:0904.4063 [hep-ex]].
\bibitem{Dungel:2010uk} 
  W.~Dungel {\it et al.}  [Belle Collaboration],
  Phys.\ Rev.\ D {\bf 82}, 112007 (2010)
  [arXiv:1010.5620 [hep-ex]].
  
\end{thebibliography}
\end{document}